\documentclass[10pt,twocolumn,twoside] {IEEEtran}
\usepackage{graphicx,amsfonts}
\usepackage{subfig,color}
\usepackage{cite}
\ifCLASSINFOpdf
\else
\fi
\usepackage[cmex10]{amsmath}
\usepackage{array,multirow}
\usepackage{mdwmath}
\usepackage{mdwtab}

\DeclareMathOperator*{\argmin}{arg\,min}
\DeclareMathOperator*{\argmax}{arg\,max}

\begin{document}
\setlength{\skip\footins}{2mm}
\addtolength{\abovedisplayskip}{-1.7mm}
\addtolength{\belowdisplayskip}{-1.2mm}
%
\title{Nonnegative HMM for Babble Noise Derived from Speech HMM: Application to Speech Enhancement}

\author{Nasser~Mohammadiha, Arne~Leijon
\thanks{
At the time of writing this paper, the authors were with the Sound and Image Processing Laboratory, School of Electrical Engineering, KTH Royal Institute of Technology. Nasser Mohammadiha is currently with Zenuity AB.}}

%
%

\markboth{This paper is published at IEEE TRANSACTIONS ON AUDIO, SPEECH, AND LANGUAGE PROCESSING, 2013}%
{Mohammadiha \MakeLowercase{\textit{et al.}}: NHMM for babble noise}
%



\maketitle

\begin{abstract}
Deriving a good model for multitalker babble noise can facilitate different speech processing algorithms, e.g. noise reduction, to reduce the so-called cocktail party difficulty. In the available systems, the fact that the babble waveform is generated as a sum of N different speech waveforms is not exploited explicitly. In this paper, first we develop a gamma hidden Markov model for power spectra of the speech signal, and then formulate it as a sparse nonnegative matrix factorization (NMF). Second, the sparse NMF is extended by relaxing the sparsity constraint, and a novel model for babble noise (gamma nonnegative HMM) is proposed in which the babble basis matrix is the same as the speech basis matrix, and only the activation factors (weights) of the basis vectors are different for the two signals over time. Finally, a noise reduction algorithm is proposed using the derived speech and babble models. All of the stationary model parameters are estimated using the expectation-maximization (EM) algorithm, whereas the time-varying parameters, i.e. the gain parameters of speech and babble signals, are estimated using a recursive EM algorithm. The objective and subjective listening evaluations show that the proposed babble model and the final noise reduction algorithm significantly outperform the conventional methods.
\end{abstract}


%
\IEEEpeerreviewmaketitle

\vspace{-3mm}
\section{Introduction}
Multispeaker babble noise is one of the frequently encountered interferences in daily life that greatly degrades the quality and intelligibility of a
target speech signal. The problem of understanding the desired speech in the presence of other interfering speech signals and background
noise (also known as the ``cocktail party problem'') has received great attention since it was popularized by Cherry in 1953 \cite{Cherry1953}.
Different auditory aspects of this problem are investigated (e.g. \cite{Arons1992,Haykin2005}), and the intelligibility of speech in the presence of multitalker babble noise is examined (e.g. \cite{Simpson2005}). In addition, there have been few studies that have addressed some
babble-specific signal processing techniques to improve speech perception in the presence of background babble noise. In \cite{Krishnamurthy2009},
considering a single-channel observation of the babble noise, a framework was proposed to characterize the underlying babble signal. Also, the
effect of the number of conversations and speakers was investigated, and a system was proposed to identify the number of speakers in
a presented babble noise; moreover, it was shown that this information is beneficial for speaker recognition.

On the contrary, little attention has been paid to develop mathematical babble-specific models that can also be used in signal processing
algorithms, e.g. speech enhancement. The goal of speech enhancement algorithms is to improve the quality and intelligibility of the noisy speech, e.g., \cite{Ephraim1984,Shen1999,Vermaak2002,Martin2005,Volodya2006,Erkelens2007}, and among different applications, it is very beneficial for hearing aid users \cite{Levitt2001}. Various classes of single channel model-based speech enhancement
approaches have been proposed in the literature. In these methods, for each type of signal (speech or noise) a model
is considered and the model parameters are obtained using the training samples of that signal. Then, the task of the speech enhancement is
done by defining an interactive model between the speech and noise signals. Some examples of this class of algorithms include the codebook-based approaches\cite{Srinivasan2006} and HMM-based methods \cite{Ephraim1992,Sameti1998,Zhao2007}. However, none of these methods exploit the fact that the babble is generated by adding different speech signals, and hence the structure of the considered model for babble noise is similar to that of other noise types. In this paper, we derive a statistical model for babble noise, which takes into account the fact that the babble is generated by adding speech signals of $M$ independent speakers. Then, we propose
a single-channel speech enhancement framework that utilizes the derived babble model to enhance the noisy speech signal.

The proposed babble model is based on the nonnegative matrix factorization (NMF). NMF is a technique to approximate a nonnegative matrix $\mathbf{X}$
by a nonnegative linear combination of some basis vectors \cite{Lee1999}, i.e. $\mathbf{X}\approx\mathbf{T}\mathbf{V}$. In speech processing:
$\mathbf{X}$ is the spectrogram of the signal with short-time spectral vectors stored as columns in $\mathbf{X}$, $\mathbf{T}$ is the basis matrix
or basis spectral vectors, and $\mathbf{V}$ is called the NMF coefficient matrix. NMF has been used successfully in different
fields including blind source separation \cite{Smaragdis2007,Fevotte2009,Ozerov2010}, and speech
enhancement \cite{wilson2008a,Mysore2011,Mohammadiha2011b,Mohammadiha2011a,Mohammadiha2012a}. The ``pure addition'' property of NMF makes it a powerful technique
to be used whenever some nonnegative quanta are added to each other. In the case of babble noise, spectral vectors of different speech
signals can be added to generate a spectral vector of babble.

The basic idea of NMF-based speech enhancement algorithms is that, for
each signal, an NMF model is considered and its parameters are obtained using the training data. Then, a mixing model is defined, which usually involves the assumption that the spectrograms of the noise and speech signals are additive, and speech enhancement is carried out by a Wiener-type
filtering approach. Two important shortcomings of NMF have to be considered when designing NMF-based speech enhancement systems:

1) The correlation between consecutive time-frames is not handled directly in a standard NMF. To overcome this problem, several approaches have been proposed
\cite{wilson2008a,Mysore2011,Mohammadiha2011a,Mohammadiha2011b,Mohammadiha2012a}. For instance, a semi-supervised approach (where the noise type is
not known a priori) was proposed in \cite{Mysore2011}, which was based on a nonnegative hidden Markov model (NHMM) where
the correlation of the signals were taken into account by the transition probability matrix of the underlying HMM. In \cite{Mohammadiha2012a},
a Bayesian NMF based speech enhancement algorithm was proposed in which the temporal correlation of the underlying speech and noise
signals was exploited through the informative prior distributions.

2) For some noise signals, the noise basis matrix is quite similar to the basis matrix
of the speech signal, e.g. the basis matrix of the babble noise should be quite similar to the basis matrix of the speech signal. As a result, the performance of the noise reduction algorithms is usually worse in the case of babble noise \cite{wilson2008a,Mohammadiha2011a}. This issue has not been addressed in the available systems and is one of the main focuses of this study.

In this paper, first we derive an ergodic gamma-HMM model for the power spectral coefficients of the speech signal. Next, we formulate
the speech model as a sparse NMF. Then, by relaxing the sparsity constraint, we derive a gamma nonnegative hidden Markov model (gamma-NHMM) for
babble noise in which the basis matrix is identical to the speech basis matrix, and only the activity of the basis vectors segregates
the speech from the babble signal. Moreover, an expectation-maximization (EM) algorithm is proposed to estimate the model parameters. In addition, to employ the derived
babble model for speech enhancement, an HMM-based speech enhancement framework in the time-frequency domain is proposed where each power
spectral vector of the power spectrogram of speech and babble signals are modeled by the gamma-HMM and gamma-NHMM models, respectively.
The proposed framework differs from the state-of-the-art HMM-based approaches \cite{Ephraim1992,Sameti1998,Zhao2007} as
we directly model the spectral vectors with HMM. In the available HMM-based methods, the waveform signal is modeled as an autoregressive
(AR) process, and hence the waveforms of speech and noise signals are modeled by HMM. Thus, this new framework facilitates a new class
of HMM-based speech enhancement algorithms. Similar to \cite{Zhao2007}, the interaction model for the noisy speech signal is constructed by
considering a prior distribution over the long-term energy levels of the speech and noise signals. A recursive EM algorithm \cite{Titterington1984,Krishnamurthy1993}
is developed to estimate the time-varying parameters of these distributions online. The excellence of the proposed babble model and noise reduction
scheme is demonstrated through objective evaluations and a subjective listening test.

The rest of the paper is organized as follows: The gamma-HMM speech signal model is developed in Section \ref{sec:Speech-Signal-Model}.
In Section \ref{sec:Babble-model}, the gamma-NHMM model of babble noise is derived. In Section \ref{sec:Enhancement-Method}, the mixed
signal model and noise reduction algorithm is constructed. The estimation of the stationary model parameters and time-varying parameters is described in
Section \ref{sec:Parameter-Estimation}. The objective and subjective examination of the noise reduction algorithms are presented in Section \ref{sec:Experiments-and-Results}. Finally, Section \ref{sec:Conclusion} concludes the study.
\vspace{-3mm}
\section{Speech Signal Model\label{sec:Speech-Signal-Model}}
\subsection{Single-voice Gamma HMM\label{sub:Single-voice-Gamma-HMM}}
We model the magnitude-squared DFT coefficients (periodogram coefficients) of the speech signal using an $\bar{N}$-state HMM with gamma distributions as output
probability density functions. Throughout this paper, random variables are represented with capital letters, e.g. $\mathbf{X}=\left[X_{kt}\right]$
denotes the matrix of random variables associated with the DFT coefficients of the clean speech, where $k$ is the frequency bin and $t$ denotes
the time-frame index. The corresponding realizations are shown with small letters, e.g. $\mathbf{x}$=$\left[x_{kt}\right]$. Also, let
$\left|\cdot\right|^{2}$ represents the element-wise magnitude-square operator. The conditional distribution of $\left|X_{kt}\right|^{2}$
is given as:
\begin{gather}
f\left(\left|x_{kt}\right|^{2}\mid\bar{S}_{t}=i,G_{t}=g_{t}\right)=\nonumber \\
\frac{\left(\left|x_{kt}\right|^{2}\right)^{\alpha_{k}-1}}{\left(g_{t}b_{ki}\right)^{\alpha_{k}}\Gamma\left(\alpha_{k}\right)}e^{-\left|x_{kt}\right|^{2}/\left(g_{t}b_{ki}\right)},
\label{eq:speech_gamma_model}
\end{gather}
where the conditional density $f_{X\mid Y}\left(x\mid Y=y\right)$ is simply shown as $f\left(x\mid Y=y\right)$ to keep notations uncluttered,
and $\Gamma\left(\cdot\right)$ is the Gamma function. Here, $\bar{S}_{t}$ is the random hidden state of the speech signal, $\alpha_{k}$ is
the shape parameter, $b_{ki}$ is the scale parameter, and $G_{t}$ is the stochastic gain parameter, which is discussed later. The expected value and variance of $X_{kt}$ are defined as: $E(\left|X_{kt}\right|^{2}\mid\bar{S}_{t}=i,G_{t}=g_{t})=\alpha_{k}g_{t}b_{ki}$,
and $\text{var}(\left|X_{kt}\right|^{2}\mid\bar{S}_{t}=i,G_{t}=g_{t})=\alpha_{k}\left(g_{t}b_{ki}\right)^{2}$.

The gamma assumption for a magnitude-squared DFT coefficient in \eqref{eq:speech_gamma_model} is motivated by the super-Gaussianity of the speech DFT coefficients \cite{Martin2005,Erkelens2007}. Denote the real and imaginary parts of the DFT coefficient by $\text{Re}\{X_{kt}\}$ and $\text{Im}\{X_{kt}\}$, respectively. Assuming that $\text{Re}\{X_{kt}\}$ and $\text{Im}\{X_{kt}\}$ have a two-sided generalized gamma distribution is equivalent to assuming that $|\text{Re}\{X_{kt}\}|$ and $|\text{Im}\{X_{kt}\}|$
have a generalized gamma distribution. Then, it can be easily shown that $|\text{Re}\{X_{kt}\}|^{2}$ and $|\text{Im}\{X_{kt}\}|^{2}$
have a gamma distribution if the $\gamma$ parameter of the generalized gamma distributions equals 2 (see \cite{Erkelens2007} for a general discussion and definition of $\gamma$).
We use the standard assumption that $\text{Re}\{X_{kt}\}$ and $\text{Im}\{X_{kt}\}$ are independent and identically distributed. Since the sum of two
independent gamma random variables (RV) with equal scale parameters is a gamma RV, $\left|X_{kt}\right|^{2}=|\text{Re}\{X_{kt}\}|^{2}+|\text{Im}\{X_{kt}\}|^{2}$
will have a gamma distribution.

In general, state-conditional densities can describe different parts of the speech signal depending on the total number of states. For example, when 50$\sim$60 states are available, each state roughly corresponds to one phoneme.

\begin{figure}[!t]
\centering
\includegraphics[scale=0.8]{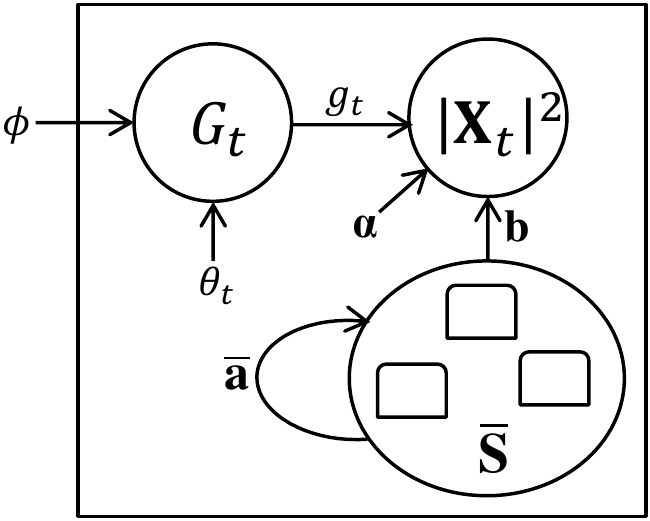}
\vspace{-1mm}
\caption{A schematic representation of the HMM with gain modeling. $\bar{\mathbf{a}}$ dentes the transition probability matrix.}
\label{fig:HMM-model}
\vspace{-7mm}
\end{figure}

The short-term stochastic gain parameter $G_t$ in \eqref{eq:speech_gamma_model} is considered to model the long-term changes in the speech energy level over time. Since $G_t$ is nonnegative, we choose to have a gamma distribution to govern $G_t$ in order to simplify the resulting algorithm:
\begin{equation}
f\left(g_{t}\right)=\frac{g_{t}^{\phi-1}}{\theta_{t}^{\phi}\Gamma\left(\phi\right)}e^{-g_{t}/\theta_{t}},
\label{eq:gain_speech}
\end{equation}
where $\phi$ and $\theta_t$ are the shape and scale parameters, respectively. In this model, the long-term speech level is modeled by the time-varying scale parameter $\theta_{t}$, while relative signal-energy levels for different states are modeled by $b_{ki}$
(see Fig.~\ref{fig:HMM-model}). Also, we have: $E\left(G_{t}\right)=\phi\theta_{t}$,
and $\text{var}\left(G_{t}\right)=\phi\theta_{t}^{2}$. Since $\sqrt{\text{var}\left(G_{t}\right)}/E\left(G_{t}\right)=\sqrt{\phi}$,
by using (\ref{eq:gain_speech}) we assume that in the log-domain the standard deviation of outcomes of $G_{t}$ from its mean value is approximately constant, independent
of the long-term level of speech. Considering that $E\left(G_{t}\right)$ is updated for different speech levels, the above assumption of gamma distribution for $G_t$ is reasonable.

The complete HMM output density functions can now be expressed as:
{\setlength\arraycolsep{-10em}
\begin{equation}
f\hspace{-.25em}\left(\left|\mathbf{x}_{t}\right|^{2}\mid\bar{S}_{t}=i,G_{t}=g_{t}\right)\hspace{-.15em}=\hspace{-.15em}\prod_{k=1}^{K}f\hspace{-.25em}\left(\left|x_{kt}\right|^{2}\mid\bar{S}_{t}=i,G_{t}=g_{t}\right),
\label{eq:vector_speech_dist}
\end{equation}
}where we have assumed that DFT coefficients at different frequency bins are conditionally independent  \cite{Ephraim1984,Martin2005,Erkelens2007}.
The state-conditional probability of the observed power spectral coefficients of the speech signal (which will be used for parameter estimation) can now be computed by integrating out the gain variable. Using the properties of the generalized inverse Gaussian distribution (see Appendix \ref{app.Posteior_dist_gain}), this can be obtained in a closed form as:
\begin{gather*}
f\hspace{-.25em}\left(\left|\mathbf{x}_{t}\right|^{2}\mid\bar{S}_{t}=i\right)=\int_{0}^{\infty}f\hspace{-.25em}\left(\left|\mathbf{x}_{t}\right|^{2}\mid\bar{S}_{t}=i,G_{t}=g_{t}\right)f\hspace{-.15em}\left(g_{t}\right)dg_{t}\\
=\frac{2\tau^{\nu/2}\mathcal{K}_{\vartheta}\left(2\sqrt{\rho\tau}\right)}{\rho^{\nu/2}\theta_{t}^{\phi}\Gamma\left(\phi\right)}\prod_{k=1}^{K}\frac{\left|x_{kt}\right|^{2\left(\alpha_{k}-1\right)}}{b_{ki}^{\alpha_{k}}\Gamma\left(\alpha_{k}\right)},
\end{gather*}
where we have defined $\rho=1/\theta_{t}$, $\nu=\phi-\sum_{k=1}^{K}\alpha_{k}$, $\tau=\sum_{k=1}^{K}\left|x_{kt}\right|^{2}b_{ki}^{-1}$, and $\mathcal{K}_{\vartheta}\left(\cdot\right)$ denotes a modified Bessel function of the second kind.

The sequence of hidden states is characterized by a first-order Markov model, with the transition probability matrix $\bar{\mathbf{a}}$,
and with the elements $\bar{a}_{ij}=f(\bar{S}_{t}=j\mid\bar{S}_{t-1}=i)$. As we are modeling speech in general, and not a specific utterance,
the state Markov chain is considered to be fully connected, and hence \textit{ergodic}, with the time-invariant state probability mass vector $\mathbf{\bar{p}}$,
and with the elements $\bar{p}_{i}=f(\bar{S}_{t}=i)$.
\vspace{-2.5mm}
\subsection{Gamma-HMM as a Probabilistic NMF \label{sub:Gamma-HMM-as}}
Instead of denoting the hidden state by its index number, as $\bar{S}_{t}=i$, we can denote the random discrete state by a one-of-$\bar{N}$ indicator
column vector $\bar{\mathbf{S}}_{t}$, where $\bar{S}_{it}=1$ and $\bar{S}_{jt}=0,j\neq i$. Using this notation,
the selected state-conditional set of scale parameters $\mathbf{b}_{i}=\left(b_{1i},\ldots,b_{Ki}\right)^{\top}$, given a particular state $\bar{\mathbf{S}}_{t}$ with $\bar{S}_{it}=1$, can
be simply expressed by $\mathbf{b\bar{\mathbf{S}}_{t}}$, where all of the $\bar{N}$ state-conditional scale parameter vectors have been collected as columns in the {}``basis'' matrix $\mathbf{b}=\left(\mathbf{b}_{1},\mathbf{b}_{2},\ldots,\mathbf{b}_{\bar{N}}\right).$

The complete sequence of scale parameter vectors for the complete spectrum sequence $\left|\mathbf{X}\right|^{2}=(\left|\mathbf{X}_{1}\right|^{2},\left|\mathbf{X}_{2}\right|^{2},\ldots,\left|\mathbf{X}_{t}\right|^{2},\ldots)$
can then be expressed as $\mathbf{b}\bar{\mathit{\mathbf{S}}}$ where $\bar{\mathit{\mathbf{S}}}=(\bar{\mathbf{S}}_{1},\bar{\mathbf{S}}_{2},\ldots,\bar{\mathbf{S}}_{t},\ldots)$
is the random sequence of the state indicator vectors. The probability of the complete sequence $\left|\mathbf{x}\right|^{2}$ of the observed
short-time spectra, given any state sequence $\bar{\mathbf{s}}$ and gain factors $\mathbf{g}=\left(g_{1},g_{2},\ldots,g_{t},\ldots\right)$,
can now be obtained as:
\begin{equation}
f\left(\left|\mathbf{x}\right|^{2}\mid\bar{\mathbf{s}},\mathbf{g}\right)=\prod_{t}f\left(\left|\mathbf{x}_{t}\right|^{2}\mid\bar{\mathbf{s}}_{t},g_{t}\right),
\label{eq:speech_NMF_model}
\end{equation}
\begin{equation}
f\hspace{-.25em}\left(\left|\mathbf{x}_{t}\right|^{2}\mid\bar{\mathbf{s}}_{t},g_{t}\right)=\prod_{k}\frac{\left(\left|x_{kt}\right|^{2}\right)^{\alpha_{k}-1}}{\left(g_{t}\left[\mathbf{b}\mathbf{\bar{\mathbf{s}}}_{t}\right]_{k}\right)^{\alpha_{k}}\Gamma\left(\alpha_{k}\right)}e^{-\left|x_{kt}\right|^{2}/\left(g_{t}\left[\mathbf{b\bar{\mathbf{s}}}_{t}\right]_{k}\right)},
\label{eq:speech_NMF_model_per_frame}
\end{equation}
where $\left[\cdot\right]_{k}$ denotes the $k^{\text{th}}$ element of the vector, and we have: $E(\left|X_{kt}\right|^{2}\mid\bar{\mathbf{s}}_{t},g_{t})=\alpha_{k}g_{t}\left[\mathbf{b}\mathbf{\bar{\mathbf{s}}}_{t}\right]_{k}$. Eq. (\ref{eq:speech_NMF_model_per_frame}) can be used to derive an NMF representation of any observed nonnegative matrix $\left|\mathbf{x}\right|^{2}$.
In order to show this, we approximate an observed sequence by its expected value, under the model assumptions, and show that the expected value is the product of two nonnegative matrices. To compute the expected value, the posterior
distribution of the state and gain variables are employed. That is, an NMF approximation $\widehat{\left|\mathbf{x}_{t}\right|^{2}}$ of an input vector $\left|\mathbf{x}_{t}\right|^{2}$ is given as:
\begin{gather}
\widehat{\left|\mathbf{x}_{t}\right|^{2}}=\nonumber \\
\sum_{\bar{\mathbf{s}}_{t}}\int E\left(\left|\mathbf{X}_{t}\right|^{2}\mid\bar{\mathbf{s}}_{t},g_{t}\right)f\left(\bar{\mathbf{s}}_{t},g_{t}\mid\left|\mathbf{x}\right|^{2}\right)dg_{t}.
\label{eq:expexted_NMF}
\end{gather}
Let us define $\hat{\mathbf{b}}$ with elements $\hat{b}_{ki}=\alpha_{k}b_{ki}$. Noting from (\ref{eq:speech_NMF_model_per_frame}) that $E(\left|\mathbf{X}_{t}\right|^{2}\mid\bar{\mathbf{s}}_{t},g_{t})=g_{t}\hat{\mathbf{b}}\bar{\mathbf{s}}_{t}$, (\ref{eq:expexted_NMF}) can be written as:
\begin{gather}
\widehat{\left|\mathbf{x}_{t}\right|^{2}}=\nonumber \\
\hat{\mathbf{b}}\sum_{\bar{\mathbf{s}}_{t}}\int g_{t}\bar{\mathbf{s}}_{t}f\left(\bar{\mathbf{s}}_{t}\mid\left|\mathbf{x}\right|^{2}\right)f\left(g_{t}\mid\bar{\mathbf{s}}_{t},\left|\mathbf{x}\right|^{2}\right)dg_{t}.
\label{eq:nmf_interperatation}\end{gather}
Here, the conditional state probabilities $f(\bar{\mathbf{s}}_{t}\mid\left|\mathbf{x}\right|^{2})$ can be calculated using the forward-backward algorithm \cite{Rabiner1989}. Since $g_t$ depends only on the current observation, $f(g_{t}\mid\bar{\mathbf{s}}_{t},\left|\mathbf{x}\right|^{2})=f(g_{t}\mid\bar{\mathbf{s}}_{t},\left|\mathbf{x}_{t}\right|^{2})$. The posterior distribution of the gain variable is a generalized inverse Gaussian distribution (this is derived in Appendix \ref{app.Posteior_dist_gain}
and will also be used in Subsection \ref{sub:Speech-Model-Training}). Thus, the required integration in (\ref{eq:nmf_interperatation}) is available in a closed form (Eq. (\ref{eq:expected_G})). Denoting $E(g_{t}\mid\bar{\mathbf{s}}_{t},\left|\mathbf{x}_{t}\right|^{2})=\int g_{t}f(g_{t}\mid\bar{\mathbf{s}}_{t},\left|\mathbf{x}_{t}\right|^{2})dg_{t}$, and  $\mathbf{u}_{t}=\sum_{\bar{\mathbf{s}}_{t}}\bar{\mathbf{s}}_{t}f(\bar{\mathbf{s}}_{t}\mid\left|\mathbf{x}\right|^{2})E(g_{t}\mid\bar{\mathbf{s}}_{t},\left|\mathbf{x}_{t}\right|^{2})$, we can write: $\widehat{\left|\mathbf{x}_{t}\right|^{2}}=\hat{\mathbf{b}}\mathbf{u}_{t}$. Hence, the proposed gamma-HMM model can be used to factorize a nonnegative
matrix $\left|\mathbf{x}\right|^{2}$ into a nonnegative basis matrix $\hat{\mathbf{b}}$ and an NMF coefficients matrix $\mathbf{u}$ as: $\left|\mathbf{x}\right|^{2}\approx\hat{\mathbf{b}}\mathbf{u}$.
In an extremely sparse case where $f(\bar{\mathbf{s}}_{t}^{\prime}\mid\left|\mathbf{x}\right|^{2})=1$ only for one state $\bar{\mathbf{s}}_{t}^{\prime}$, depending on
time $t$, and all of the other states have a zero probability, this model reduces to: $\left|\mathbf{x}\right|^{2}\approx\hat{\mathbf{b}}\mathbf{u}^{\prime}$
with $\mathbf{u}_{t}^{\prime}=\bar{\mathbf{s}}_{t}^{\prime}E(g_{t}\mid\bar{\mathbf{s}}_{t}^{\prime},\left|\mathbf{x}_t\right|^{2})$.

\vspace{-2mm}
\section{Probabilistic Model of Babble Noise\label{sec:Babble-model}}
We model the waveform of the babble noise as a weighted sum of $M$ i.i.d. clean speech sources. Therefore, the expected value of the short-time
power spectrum vector (periodogram) of babble at time $t$, $\left|\mathbf{V}_{t}\right|^{2}$, is given by:
\begin{equation}
E\left(\left|\mathbf{V}_{t}\right|^{2}\right)=\sum_{m=1}^{M}E\left(\left|\mathbf{X}_{mt}\right|^{2}\right),
\label{eq:babble_model}
\end{equation}
where each random vector $\left|\mathbf{X}_{mt}\right|^{2}$ is independently generated by an instance of the gamma-HMM described
in Section \ref{sec:Speech-Signal-Model}. Note that in \eqref{eq:babble_model} different weights are used for different speakers as a consequence of the gain modeling in Section \ref{sec:Speech-Signal-Model}. That is, there is a hidden speaker-dependent gain ($g_{mt}$) in \eqref{eq:babble_model} (see also \eqref{eq:speech_gamma_model}). Eq \eqref{eq:babble_model} provides a simplified model of real-life babble noise because we are not modeling reverberations here. There might also be additional noise with a recorded babble, which is not considered in \eqref{eq:babble_model}. However, it must be mentioned that all of the babble model parameters will be estimated given a babble training data set, with no information about $M$. Therefore, the estimated parameters will be such that the model explains the considered babble as well as possible.

In this view, the babble noise is still described by an HMM with discrete states defined by the combination of the states for each of the $M$
sources. Since the speech signal has $\bar{N}$ states, there are $\bar{N}^{M}$ possible discrete states for the babble. As the discrete
HMM for the clean speech is already an approximation, and speech should probably rather be modeled with a continuous-state HMM, it would be
preferable to describe the babble sequence with a continuous-state HMM. On the other hand, an exact implementation of the EM algorithm for HMMs
with a continuous-state is generally not possible, except for some very few specific cases, e.g. Gaussian linear state-space models, and
simulation-based methods have to be used instead \cite{Cappe2005}. Hence, it would be preferable to avoid a continuous-state structure
whenever it is possible. Furthermore, in a real babble noise only a finite number of states (say representative states) would be
sufficient for practical purposes to model the normalized spectral shape of the signal; this is indicated, for example, by the success of the vector quantization techniques
to quantize continuous signals with a limited number of centroid vectors effectively. Based on these reasons, in the following we
model the babble noise with a discrete-state HMM.

Using the text following (\ref{eq:speech_NMF_model_per_frame}), (\ref{eq:babble_model}) can be written as:
\begin{eqnarray}
 & E\left(\left|\mathbf{V}_{t}\right|^{2}\mid\bar{\mathbf{s}}_{1t},g_{1t},\ldots\bar{\mathbf{s}}_{Mt},g_{Mt}\right)=\nonumber \\
 & \sum_{m=1}^{M}E\left(\left|\mathbf{X}_{mt}\right|^{2}\mid\bar{\mathbf{s}}_{mt},g_{mt}\right)=\nonumber \\
 & \sum_{m=1}^{M}g_{mt}\mathbf{\hat{b}}\bar{\mathbf{s}}_{mt}=\mathbf{\hat{b}}\sum_{m=1}^{M}g_{mt}\bar{\mathbf{s}}_{mt},
 \label{eq:babble_EV}
 \end{eqnarray}
where $\hat{b}_{ki}=\alpha_{k}b_{ki}$ as before.  In the babble HMM, we will now approximate the sum over $m$ in \eqref{eq:babble_EV} by the babble hidden state vectors and the gain variables. Let us denote the babble hidden state vector at time $t$ by $\ddot{\mathbf{S}}_{t}$ (as opposed
to the speech state indicator shown by $\mathbf{\bar{\mathbf{S}}}_{t}$) and its realizations by $\ddot{\mathbf{s}}_{t}$ that can take one of the $\ddot{N}$
possible state value vectors $\{\ddot{\mathbf{s}}_{1}^{\prime},\ddot{\mathbf{s}}_{2}^{\prime},\ldots\ddot{\mathbf{s}}_{\ddot{N}}^{\prime}\}$. Note that $\ddot{N}$ is different from the number of speakers in the babble, which is shown by $M$ in \eqref{eq:babble_EV}. Also, denote the stochastic babble gain by random variable $H_{t}$ and its realizations by $h_{t}$.

The power spectrum values of the babble, as defined by (\ref{eq:babble_model}), \eqref{eq:babble_EV} are not exactly gamma-distributed%
\footnote{Eq. \eqref{eq:babble_EV} is defined for the expected values; to obtain the exact distribution of the babble power spectral vectors,
given the hidden states for all of the speakers, both the summation of individual gamma distributions and the distribution of the cross terms
have to be considered.%
}, given the hidden state. However, our informal simulations showed that the babble DFT coefficients also have super-Gaussian distributions.
This is understandable, considering the similarity of speech and babble.
Hence, the same argument that was used for the speech model in Subsection \ref{sub:Single-voice-Gamma-HMM} can be used here to motivate that gamma distribution is a good approximation for the distribution of the babble spectra. We now extend the clean-speech model in (\ref{eq:speech_NMF_model_per_frame}), just slightly, to model
the density of the babble short-time power spectrum as:
\begin{equation}
f\hspace{-.25em}\left(\left|\mathbf{v}_{t}\right|^{2}\mid\ddot{\mathbf{s}}_{t},h_{t}\right)\hspace{-.15em}=\hspace{-.15em}\prod_{k}\frac{\left(\left|v_{kt}\right|^{2}\right)^{\beta_{k}-1}}{\left(h_{t}\left[\mathbf{b}\ddot{\mathbf{s}}_{t}\right]_{k}\right)^{\beta_{k}}\Gamma\left(\beta_{k}\right)}e^{-\left|v_{kt}\right|^{2}/\left(h_{t}\left[\mathbf{b}\ddot{\mathbf{s}}_{t}\right]_{k}\right)},
\label{eq:babble_NMF_model}
\end{equation}
here, the main new feature is that the hidden state vectors $\ddot{\mathbf{s}}$ are not indicator vectors (columns of $\mathbf{\bar{\mathbf{s}}}$
were one-of-$\bar{N}$ indicator vectors in (\ref{eq:speech_NMF_model_per_frame})). This is a result of (\ref{eq:babble_model}). More specifically,
if we set $\beta_{k}=\alpha_{k}$ and $\ddot{\mathbf{s}}_{t}=\sum_{m}g_{mt}\bar{\mathbf{s}}_{mt}$, then (\ref{eq:babble_NMF_model}) leads to the same expected value
as in (\ref{eq:babble_EV}) with $h_{t}=1$. In this context, $\ddot{\mathbf{s}}_t$ is the weighted sum of the $M$ indicator vectors. The shape parameters $\beta_{k}$ are still assumed to be independent of the hidden states, but may be different
%
from the shape parameters $\alpha_{k}$ of the clean speech model. In this approach, the babble signal is generated as a weighted sum of different
clean speech waveforms, thus, the {}``basis'' matrix $\mathbf{b}$ is assumed to be the same and only the weights of the basis vectors are
different for the speech and the babble signals. The short-term stochastic gain $H_{t}$ in (\ref{eq:babble_NMF_model}) is assumed to have a gamma distribution
as:
\begin{equation}
f\left(h_t\right)=\frac{h_{t}^{\psi-1}}{\gamma_{t}^{\psi}\Gamma\left(\psi\right)}e^{-h_{t}/\gamma_{t}}.
\label{eq:gain_babble}
\end{equation}
In \eqref{eq:gain_babble}, the scale parameter $\gamma_{t}$ represents the long-term energy level of the babble signal, and $\psi$ is the shape parameter. An EM-based algorithm is proposed in Subsection
\ref{sub:Noise-Training} to estimate $\ddot{N}$ babble state value vectors, $\ddot{\mathbf{s}}_{i}^{\prime}\text{ for}\ i=1,\ldots\ddot{N}$,
the state transition probabilities $\ddot{a}_{ij}=f(\ddot{\mathbf{S}}_{t}=\ddot{\mathbf{s}}_{j}^{\prime}\mid\ddot{\mathbf{S}}_{t-1}=\ddot{\mathbf{s}}_{i}^{\prime}),\beta_{k},\psi$,
and $\gamma_{t}$ given the recorded babble noise. Eq. (\ref{eq:babble_NMF_model}) is referred as gamma-NHMM since the described model performs an NMF
on the scale parameters of the HMM output distributions, which are gamma distributions.
\vspace{-2mm}
\section{Speech Enhancement Method\label{sec:Enhancement-Method}}
A noise reduction scheme is proposed in this section to enhance the speech signal that is degraded by the babble noise. The mixed signal model
is described in Subsection \ref{sub:Clean-Speech-Mixed}, which is used later in Subsection \ref{sub:Clean-Speech-Estimator} to derive an MMSE
estimator for the speech signal.

In the proposed models, the power spectra of the clean speech and the babble noise are conditionally gamma-distributed. Even though a gamma distribution might be a good approximation for the power spectra of the mixed signal, obtaining the MMSE estimator for the clean speech signal is difficult for this case (\cite{Mohammadiha2013a} proposes a solution uing this approximation). Therefore, in this part of the paper (which provides an application of the developed babble model) we limit the models to use exponential distributions for the speech and the babble power spectral coefficients ($\alpha_k=1$ in \eqref{eq:speech_gamma_model}, $\beta_k=1$ in \eqref{eq:babble_NMF_model}). This corresponds to the assumption that speech and babble DFT coefficients have complex Gaussian distributions, which have been used successfully in the literature (e.g. \cite{Ephraim1984,Cohen2006}). In the experimental section, we show that even with this additional simplification the proposed noise reduction method outperforms the competing algorithms. To keep the generality of the speech and babble models for potential future applications, the proposed parameter estimation algorithm in Section \ref{sec:Parameter-Estimation} will be given for the general gamma case.
\vspace{-3mm}
\subsection{Clean Speech Mixed with Babble\label{sub:Clean-Speech-Mixed}}
Assuming that the DFT coefficients of the clean speech and babble noise are
complex Gaussian, DFT coefficients of the mixed signal $\mathbf{Y}$,
\[
\mathbf{Y}_{t}=\mathbf{X}_{t}+\mathbf{V}_{t},\]
will also have complex Gaussian distribution. Let us represent
the composite state of the mixed signal by $\mathbf{S}_t$ that can take one of the $N=\bar{N}\ddot{N}$ possible outcomes. Defining
$\sigma_{Y_{kt}}^{2}=E(\left|Y_{kt}\right|^{2}\mid{\mathbf{s}}_{t},g_{t},h_{t})=E(\left|X_{kt}\right|^{2}\mid\bar{\mathbf{s}}_{t},g_{t})+E(\left|V_{kt}\right|^{2}\mid\ddot{\mathbf{s}}_{t},h_{t})$
we have:
\begin{equation}
f\left(y_{kt}\mid g_{t},h_{t},\mathbf{\mathbf{s}}_t\right)=\frac{1}{\pi\sigma_{Y_{kt}}^{2}}e^{-\frac{\left|y_{kt}\right|^{2}}{\sigma_{Y_{kt}}^{2}}},
\label{eq:conditiona_pdf_noisy}
\end{equation}
and also
\begin{equation}
f\left(\mathbf{y}_{t}\mid g_{t},h_{t},\mathbf{\mathbf{s}}_t\right)=\prod_{k=1}^{K}f\left(y_{kt}\mid g_{t},h_{t},\mathbf{\mathbf{s}}_t\right).
\label{eq:conditional_pdf_noisy_vector}
\end{equation}
The state-conditional distribution of the mixed signal can be obtained by integrating out the gain variables as:
\begin{equation}
f\left(\mathbf{y}_{t}\mid\mathbf{s}_{t}\right)=\int\int f\left(\mathbf{y}_{t}\mid g_{t},h_{t},\mathbf{s}_{t}\right)f\left(g_{t},h_{t}\right)dg_{t}dh_{t}.
\label{eq:true_likelihood}
\end{equation}
The required expectations to calculate $\sigma_{Y_{kt}}^{2}$ in (\ref{eq:conditiona_pdf_noisy}) are obtained considering the models given in (\ref{eq:speech_NMF_model_per_frame})
and (\ref{eq:babble_NMF_model}). The analytical evaluation of (\ref{eq:true_likelihood}) turns out to be difficult; although numerical methods can be used
to calculate the required integrations, we approximate the integrand by its behavior near to its maximum by applying Laplace approximation
\cite[Sec. 4.4]{bishop2006}. Hence, we first derive an EM algorithm to obtain the state-dependent Maximum a-Posteriori estimates $g_{t}^{\prime}$
and $h_{t}^{\prime}$ in Appendix \ref{sec:MAP-Estimate-of} based on the following optimization problem:
\begin{equation}
\left\{ g_{t}^{\prime},h_{t}^{\prime}\right\} =\argmax_{g_{t},h_{t}}\, f\left(\mathbf{y}_{t}\mid g_{t},h_{t},\mathbf{s}_{t}\right)f\left(g_{t},h_{t}\right),
\label{eq:max_for_gains}
\end{equation}
then (\ref{eq:true_likelihood}) is approximated by
\begin{equation}
f\left(\mathbf{y}_{t}\mid\mathbf{\mathbf{s}}_t\right)\approx
f\left(\mathbf{y}_{t}\mid g_{t}^{\prime},h_{t}^{\prime},\mathbf{s}_{t}\right)f\left(g_{t}^{\prime},h_{t}^{\prime}\right)\frac{2\pi}{\sqrt{\text{det}\left(A_{\mathbf{s}_{t}}\right)}},
\label{eq:Lapplace_app}
\end{equation}
where $\text{det}\left(A_{\mathbf{s}_{t}}\right)$ is the determinant of the negative Hessian of $\ln\, f\left(\mathbf{y}_{t},g_{t},h_{t}\mid\mathbf{s}_{t}\right)$
with respect to $g_{t},h_{t}$, evaluated at the maximum point. The expression for the Hessian matrix is also given in Appendix \ref{sec:MAP-Estimate-of}.
%
It should also be mentioned that in \cite{Zhao2007} an EM algorithm was developed to find the mode of the joint distribution and then
$f\left(y\mid\mathbf{s}_{t}\right)$ was approximated by $f\left(y,g_{t}^{\prime},h_{t}^{\prime}\mid\mathbf{s}_{t}\right).$
\vspace{-2.5mm}
\subsection{Clean Speech Estimator \label{sub:Clean-Speech-Estimator}}
\vspace{-.5mm}
The posterior distribution of the clean speech DFT coefficients given the noisy observations can be written as \cite{Ephraim1992,Zhao2007}:
\begin{equation}
f\left(\mathbf{x}_{t}\mid\mathbf{y}_{1}^{t}\right)=\frac{\sum_{\mathbf{s}_{t}}\mathbf{\eta}_{t}\left(\mathbf{s}_{t}\right)f\left(\mathbf{x}_{t},\mathbf{y}_{t}\mid\mathbf{s}_{t}\right)}{f\left(\mathbf{y}_{t}\mid\mathbf{y}_{1}^{t-1}\right)},
\label{eq:speech_post_dist}
\end{equation}
where $\mathbf{y}_{t1}^{t2}\hspace{-.3em}=\hspace{-.3em}\left\{\mathbf{y}_{t1},\mathbf{y}_{t1+1},\ldots,\mathbf{y}_{t2}\right\}$, and $\mathbf{\eta}_{t}\left(\mathbf{s}_{t}\right)\hspace{-.3em}=\hspace{-.3em}$ $f\left(\mathbf{s}_{t}\mid\mathbf{y}_{1}^{t-1}\right)$ is the probability of being in the composite state $\mathbf{s}_{t}$ at time $t$ given all of the noisy observations until time $t-1$, and is calculated as:
\begin{equation}
\mathbf{\eta}_{t}\left(\mathbf{s}_{t}\right)=f\left(\mathbf{s}_{t}\mid\mathbf{y}_{1}^{t-1}\right)=\sum_{\mathbf{s}_{t-1}}a_{\mathbf{s}_{t-1},\mathbf{s}_{t}}f\left(\mathbf{s}_{t-1}\mid\mathbf{y}_{1}^{t-1}\right),
\label{eq:eta_def}
\end{equation}
with $a_{\mathbf{s}_{t-1},\mathbf{s}_{t}}\hspace{-.3em}=\hspace{-.3em}f\hspace{-.1em}\left(\mathbf{S}_{t}=\mathbf{s}_{t}\mid\hspace{-.2em}\mathbf{S}_{t-1}\hspace{-.3em}=\hspace{-.3em}\mathbf{s}_{t-1}\right)=\bar{a}_{\mathbf{\bar{s}}_{t-1},\mathbf{\bar{s}}_{t}}\ddot{a}_{\mathbf{\ddot{s}}_{t-1},\mathbf{\ddot{s}}_{t}}$
because of the independency of the speech and the noise Markov chains, and $f(\mathbf{s}_{t-1}\mid\mathbf{y}_{1}^{t-1})$ is the
scaled forward variable obtained using the forward algorithm \cite{Rabiner1989}. The joint distribution of $\mathbf{X}_{t}$ and $\mathbf{Y}_{t}$ can also be written as:
\begin{gather}
f\left(\mathbf{x}_{t},\mathbf{y}_{t}\mid\mathbf{s}_{t}\right)=\int\int f\left(\mathbf{x}_{t},\mathbf{y}_{t},g_{t},h_{t}\mid\mathbf{s}_{t}\right)dg_{t}dh_{t}\approx\nonumber \\
f\left(\mathbf{x}_{t}\mid\mathbf{y}_{t},g_{t}^{\prime},h_{t}^{\prime},\mathbf{s}_{t}\right)\int\int f\left(\mathbf{y}_{t},g_{t},h_{t}\mid\mathbf{s}_{t}\right)dg_{t}dh_{t}\approx\nonumber\\
f\left(\mathbf{x}_{t}\mid\mathbf{y}_{t},g_{t}^{\prime},h_{t}^{\prime},\mathbf{s}_{t}\right)f\left(\mathbf{y}_{t},g_{t}^{\prime},h_{t}^{\prime}\mid\mathbf{s}_{t}\right)\frac{2\pi}{\sqrt{\text{det}\left(A_{\mathbf{s}_{t}}\right)}},
\label{eq:app_joint}
\end{gather}
where the second line is obtained using a point approximation for $f(\mathbf{x}_{t}\mid\mathbf{y}_{t},g_{t},h_{t},\mathbf{s}_{t})$ (similarly to \cite{Zhao2007}), and the last line is obtained by using approximation (\ref{eq:Lapplace_app}). We can also write:
\begin{eqnarray}
f\left(\mathbf{y}_{t}\mid\mathbf{y}_{1}^{t-1}\right) & = & \sum_{\mathbf{s}_{t}}\int\mathbf{\eta}_{t}\left(\mathbf{s}_{t}\right)f\left(\mathbf{x}_{t},\mathbf{y}_{t}\mid\mathbf{s}_{t}\right)d\mathbf{x}_{t}\label{eq:observation_likelihood}\\
 & \approx & \sum_{\mathbf{s}_{t}}\mathbf{\eta}_{t}\left(\mathbf{s}_{t}\right)f\left(\mathbf{y}_{t},g_{t}^{\prime},h_{t}^{\prime}\mid\mathbf{s}_{t}\right)\frac{2\pi}{\sqrt{\text{det}\left(A_{\mathbf{s}_{t}}\right)}}.\nonumber \end{eqnarray}
Denoting $\mathbf{\zeta}_{t}\left(\mathbf{s}_{t},\mathbf{y}_{t}\right)=\mathbf{\eta}_{t}\left(\mathbf{s}_{t}\right)f\left(\mathbf{y}_{t},g_{t}^{\prime},h_{t}^{\prime}\mid\mathbf{s}_{t}\right)\frac{2\pi}{\sqrt{\text{det}(A_{\mathbf{s}_t})}}$, and using (\ref{eq:app_joint}) and (\ref{eq:observation_likelihood}),
(\ref{eq:speech_post_dist}) can be written as:
\begin{equation}
f\left(\mathbf{x}_{t}\mid\mathbf{y}_{1}^{t}\right)=\frac{\sum_{\mathbf{s}_{t}}\mathbf{\zeta}_{t}\left(\mathbf{s}_{t},\mathbf{y}_{t}\right)f\left(\mathbf{x}_{t}\mid\mathbf{y}_{t},g_{t}^{\prime},h_{t}^{\prime},\mathbf{s}_{t}\right)}{\sum_{\mathbf{s}_{t}}\mathbf{\zeta}_{t}\left(\mathbf{s}_{t},\mathbf{y}_{t}\right)}.
\label{eq:speec_post_dist_2}
\end{equation}
Because of the Gaussian assumption, calculating the state-conditional posterior distribution of the clean speech DFT coefficients is straightforward
and is given by a complex Gaussian distribution with the mean value obtained via the Wiener filtering:
\begin{equation}
E\left(X_{kt}\mid\mathbf{y}_{t},g_{t}^{\prime},h_{t}^{\prime},\mathbf{s}_{t}\right)=C_{X_{kt}}\left(C_{X_{kt}}+C_{V_{kt}}\right)^{-1}y_{kt},
\label{eq:wiener_state}
\end{equation}
and the covariance matrix given as:
\begin{gather}
E\left(\left|X_{kt}-E\left(X_{kt}\mid\mathbf{y}_{t},g_{t}^{\prime},h_{t}^{\prime},\mathbf{s}_{t}\right)\right|^{2}\mid\mathbf{y}_{t},g_{t}^{\prime},h_{t}^{\prime},\mathbf{s}_{t}\right)\nonumber \\
=C_{X_{kt}}-C_{X_{kt}}\left(C_{X_{kt}}+C_{V_{kt}}\right)^{-1}C_{X_{kt}},
\label{eq:wiener_covariance}
\end{gather}
where $C_{X_{kt}}=E(\left|X_{kt}\right|^{2}\mid g_{t}^{\prime},\mathbf{s}_{t})=\alpha_{k}g_{t}^{\prime}\left[\mathbf{b}\mathbf{\bar{\mathbf{s}}}_{t}\right]_{k}$
and $C_{V_{kt}}=E(\left|V_{kt}\right|^{2}\mid h_{t}^{\prime},\mathbf{s}_{t})=\beta_{k}h_{t}^{\prime}\left[\mathbf{b}\ddot{\mathbf{s}}_{t}\right]_{k}$.
By using (\ref{eq:wiener_state}) and (\ref{eq:speec_post_dist_2}), the MMSE estimator of the clean speech DFT coefficients is derived
as:
\begin{eqnarray}
\mathbf{\hat{x}}_{t} & = & E\left(\mathbf{X}_{t}\mid\mathbf{y}_{1}^{t}\right)\nonumber \\
 & = & \frac{\sum_{\mathbf{s}_{t}}\mathbf{\zeta}_{t}\left(\mathbf{s}_{t},\mathbf{y}_{t}\right)E\left(\mathbf{X}_{t}\mid\mathbf{y}_{t},g_{t}^{\prime},h_{t}^{\prime},\mathbf{s}_{t}\right)}{\sum_{\mathbf{s}_{t}}\mathbf{\zeta}_{t}\left(\mathbf{s}_{t},\mathbf{y}_{t}\right)},
 \label{eq:final_estimate_speech_com}
 \end{eqnarray}
or equivalently as $\hat{x}_{kt}=\kappa_{kt}y_{kt}$ where the gain parameter is given by:
\begin{equation}
\kappa_{kt}=\frac{\sum_{\mathbf{s}_{t}}\mathbf{\zeta}_{t}\left(\mathbf{s}_{t},\mathbf{y}_{t}\right)C_{X_{kt}}\left(C_{X_{kt}}+C_{V_{kt}}\right)^{-1}}{\sum_{\mathbf{s}_{t}}\mathbf{\zeta}_{t}\left(\mathbf{s}_{t},\mathbf{y}_{t}\right)}.
\label{eq:final_gain_se}
\end{equation}
\section{Parameter Estimation\label{sec:Parameter-Estimation}}
\subsection{Speech Model Training\label{sub:Speech-Model-Training}}
The EM-based Baum-Welch algorithm is followed to train speech and noise models \cite{Rabiner1989,Bilmes1997}. The parameters of the
speech model are denoted by $\lambda=\left\{\bar{\mathbf{a}},\mathbf{b},\boldsymbol{\alpha},\phi,\theta\right\} .$
Letting the training data consist of $R$ speech utterances, it is assumed that the time-dependent scale parameter of the stochastic
gain, $\theta$, remains constant during each utterance for simplicity, hence, denoted by $\theta_{r}$ in the following.

Denote the whole training set by $\bar{\mathbf{o}}=\left\{\left(\bar{\mathbf{o}}_{1,1}\ \ldots\bar{\mathbf{o}}_{1,T_{1}}\right),\ldots\left(\bar{\mathbf{o}}_{R,1}\ \ldots\bar{\mathbf{o}}_{R,T_{R}}\right)\right\} $ where $\bar{\mathbf{o}}_{r,t}=\left[\bar{o}_{r,kt}\right]$ represents
the speech power spectral vector of the $r^{\text{th}}$ sentence at time $t$. Similarly, let $\mathbf{\bar{Z}}=\{\mathbf{\bar{S}},\mathbf{G}\}$
represent the hidden variables in the model, which are not observed. Then, the maximization step in the EM algorithm consists of maximizing
\begin{equation}
Q\left(\hat{\lambda},\lambda\right)=\sum_{\mathbf{\bar{s}}}\int f\left(\bar{\mathbf{z}}\mid\bar{\mathbf{o}},\lambda\right)\ln\left(f\left(\bar{\mathbf{z}},\bar{\mathbf{o}}\mid\hat{\lambda}\right)\right)d\mathbf{g},
\label{eq:EM_cost}
\end{equation}
w.r.t $\hat{\lambda}$, where $\lambda$ is the estimated parameters from the previous iteration of the EM algorithm. $Q(\hat{\lambda},\lambda)$
can be written as:
\small
\begin{gather}
Q\left(\hat{\lambda},\lambda\right)=\hat{Q}\left(\hat{\lambda},\lambda\right)+\sum_{r,t,i}\omega_{t,r}\left(i\right)\int f\left(g_{r,t}\mid\bar{\mathbf{o}}_{r,t},\bar{S}_{r,t}=i,\lambda\right)\cdot\nonumber \\
\left(\ln f\left(g_{r,t}\mid\hat{\lambda}\right)+\ln f\left(\bar{\mathbf{o}}_{r,t}\mid g_{r,t},\bar{S}_{r,t}=i,\hat{\lambda}\right)\right)dg_{r,t},
\label{eq:cost_total_train_speech}
\end{gather}
\normalsize for $r=1\ldots R$, $t=1\ldots T_{r}$, and $i=1\ldots\bar{N}$. Here, $\hat{Q}(\hat{\lambda},\lambda)$ includes the terms
for optimizing the transition matrix $\hat{\bar{\mathbf{a}}}$ and is maximized using the standard Baum-Welch algorithm. The posterior state probabilities
\[
\omega_{t,r}\left(i\right)=f\left(\bar{S}_{r,t}=i\mid\bar{\mathbf{o}},\lambda\right),\] are obtained by the forward-backward algorithm \cite{Rabiner1989}.
To obtain the new parameters, (\ref{eq:cost_total_train_speech}) is differentiated w.r.t the parameters of interest, and the result
is set to zero. The objective function in (\ref{eq:cost_total_train_speech}) is separable for on the one hand $\{ \hat{\mathbf{b}},\hat{\boldsymbol{\alpha}}\}$, and on the other hand $\{\hat{\phi},\hat{\theta}\}$. First, consider $\{ \hat{\mathbf{b}},\hat{\boldsymbol{\alpha}}\} $; obtaining
the gradient w.r.t. $\hat{b}_{ki}$ and setting it to zero yields the following estimate:
\begin{equation}
\hat{b}_{ki}=\frac{\sum_{r,t}\omega_{t,r}\left(i\right)\bar{o}_{r,kt}E_{G_{r,t}\mid\bar{S}_{r,t},\lambda}\left(G_{r,t}^{-1}\right)}{\hat{\alpha}_{k}\sum_{r,t}\omega_{t,r}\left(i\right)}=\frac{\mu_{ki}^{o}}{\hat{\alpha}_{k}},
\label{eq:update_scale_train_speech}
\end{equation}
where $E_{G_{r,t}\mid\bar{S}_{r,t},\lambda}\left(\cdot\right)$ is the expectation w.r.t. the posterior distribution of the gain variable
$f\left(g_{r,t}\mid\bar{\mathbf{o}}_{r,t},\bar{S}_{r,t}=i,\lambda\right)$, and $\mu_{ki}^{o}$ is defined as:
\begin{equation}
\mu_{ki}^{o}=\frac{\sum_{r,t}\omega_{t,r}\left(i\right)\bar{o}_{r,kt}E_{G_{r,t}\mid\bar{S}_{r,t},\lambda}\left(G_{r,t}^{-1}\right)}{\sum_{r,t}\omega_{t,r}\left(i\right)}.
\label{eq:mu_ki}
\end{equation}
Inserting (\ref{eq:update_scale_train_speech}) into (\ref{eq:cost_total_train_speech}), and setting the gradient of the objective function w.r.t. $\hat{\alpha}_{k}$
to zero yields:
\begin{gather}
\varphi\left(\hat{\alpha}_{k}\right)-\ln\left(\hat{\alpha}_{k}\right)=\frac{1}{\sum_{r,t,i}\omega_{t,r}\left(i\right)}\times\nonumber \\
\sum_{r,t,i}\omega_{t,r}\left(i\right)\left(\ln\bar{o}_{r,kt}-E_{G_{r,t}\mid\bar{S}_{r,t},\lambda}\left(\ln G_{r,t}\right)-\ln\mu_{ki}^{o}\right),
\label{eq:update_shape_train_speech}
\end{gather}
where $\varphi\left(u\right)=\frac{d}{du}\ln\Gamma\left(u\right)$ is the digamma function, and $\mu_{ki}^{o}$ is defined in (\ref{eq:mu_ki}).
Hence, (\ref{eq:update_shape_train_speech}) is solved first, e.g. by Newton's method, and the obtained $\hat{\alpha}_{k}$ is inserted
into (\ref{eq:update_scale_train_speech}) to estimate $\hat{b}_{ki}$. Similarly, $\hat{\phi}$ and $\hat{\theta}$ can be obtained
by first estimating the shape parameter $\phi$ as:
\begin{gather}
\varphi\left(\hat{\phi}\right)-\ln\left(\hat{\phi}\right)=\frac{1}{\sum_{t,r,i}\omega_{t,r}\left(i\right)}\times\nonumber \\
\sum_{t,r,i}\omega_{t,r}\left(i\right)\left(E_{G_{r,t}\mid\bar{S}_{r,t},\lambda}\left(\ln G_{r,t}\right)-\ln\mu_{r}^{g}\right),
\label{eq:update_shape_gain_train_speech}
\end{gather}
with $\mu_{r}^{g}$ defined as:
\[
\mu_{r}^{g}=\frac{\sum_{t,i}\omega_{t,r}\left(i\right)E_{G_{r,t}\mid\bar{S}_{r,t},\lambda}\left(G_{r,t}\right)}{\sum_{t,i}\omega_{t,r}\left(i\right)},\]
and then using $\hat{\phi}$ to obtain $\hat{\theta}_{r}=\mu_{r}^{g}/\hat{\phi}$.
Since the gamma probability density function given in (\ref{eq:speech_gamma_model}) is log concave in $\alpha_{k}$ and $b_{ki}$ around the stationary
points $\hat{b}_{ki},\hat{\alpha}_k$, these update rules are guaranteed to increase the overall log likelihood score of the parameters
\cite{Levinson1986}. To perform the updates, it is required to calculate the posterior expected values of the functions of the gain variables.
The posterior distribution of the gain variable, $f\left(g_{r,t}\mid\bar{\mathbf{o}}_{r,t},\bar{S}_{r,t}=i,\lambda\right)$, is obtained in Appendix \ref{app.Posteior_dist_gain} and is a generalized inverse Gaussian. The required expected values $E_{G_{r,t}\mid\bar{S}_{r,t},\lambda}\left(G_{r,t}\right)$,
$E_{G_{r,t}\mid\bar{S}_{r,t},\lambda}\left(G_{r,t}^{-1}\right)$, and $E_{G_{r,t}\mid\bar{S}_{r,t},\lambda}\left(\ln G_{r,t}\right)$
are given with (\ref{eq:expected_G}), (\ref{eq:expected_G-1}), and (\ref{eq:expected_lnG}), respectively.
\vspace{-4mm}
\subsection{Babble Model Training\label{sub:Noise-Training}}
\vspace{-1mm}
The parameters of the babble model are denoted by $\lambda=\left\{ \ddot{\mathbf{a}},\ddot{\mathbf{s}}^{\prime},\boldsymbol{\beta},\psi,\gamma\right\} $
where $\ddot{\mathbf{s}}^{\prime}=\{ \ddot{\mathbf{s}}_{1}^{\prime},\ddot{\mathbf{s}}_{2}^{\prime},\ldots\ddot{\mathbf{s}}_{\ddot{N}}^{\prime}\} $
is the set of the $\ddot{N}$ babble state value vectors. These vectors are in principle the weighting factors associated with the basis
matrix $\mathbf{b}$. Letting the training data consist of $R$ different recordings of babble noise, similar to the speech model, it is assumed that the time-dependent
scale parameter of the stochastic gain $H$ remains constant during each recording, hence, denoted by $\gamma_{r}$ in the following.

Denote the whole training set by $\ddot{\mathbf{o}}$.
The main difference between noise
training and speech training is that for the babble training we must also update the babble state value vectors $\ddot{\mathbf{s}}_{i}^{\prime}$
for $i=1,\ldots\ddot{N}$ simultaneously with the other parameters. The update rules for $\ddot{\mathbf{a}},\gamma$, and $\psi$ are similar to
the update rules of $\bar{\mathbf{a}},\theta$, and $\phi$, respectively. The estimation of $\boldsymbol{\beta}$ and $\ddot{\mathbf{s}}^{\prime}$ are coupled. Hence, it is easier
to optimize the EM help function $Q(\hat{\lambda},\lambda)$ w.r.t. these parameters separately, given the previous estimates of
them. Obtaining the derivative of $Q(\hat{\lambda},\lambda)$ w.r.t. $\beta_k$ and setting it to zero yields the following
estimate for $\beta_k$:
\begin{gather}
\varphi\left(\hat{\beta}_{k}\right)=\frac{\sum_{r,t,i}\omega_{t,r}\left(\ddot{\mathbf{s}}_{i}^{\prime}\right)}{\sum_{r,t,i}\omega_{t,r}\left(\ddot{\mathbf{s}}_{i}^{\prime}\right)}\times\nonumber \\
\left(\ln\ddot{o}_{r,kt}-\ln\left[\mathbf{b}\ddot{\mathbf{s}}_{i}^{\prime}\right]_{k}-E_{H_{r,t}\mid\ddot{\mathbf{S}}_{r,t},\lambda}\left(\ln H_{r,t}\right)\right).
\label{eq:update_shape_train_noise}
\end{gather}
The update rule for $\ddot{\mathbf{s}}_{i}^{\prime}$ cannot be obtained in a closed form. Here, we present an approach based on the concave-convex
procedure (CCCP) \cite{Yuille2003,Sriperumbudur2009} to iteratively maximize the EM help function $Q(\hat{\lambda},\lambda)$.
CCCP is a procedure to find a local minimum of a nonconvex function and is often used to minimize a cost function that can be written
as a difference of convex functions. The core idea of this procedure is that the nonconvex function is approximated by a convex function, which can be easily minimized, and then the procedure is iterated until a local minimum is found. The negative EM help function for the babble model is given as:
\begin{gather}
-Q\left(\hat{\lambda},\lambda\right)=Q'+\nonumber \\
\sum_{t,r,i}\omega_{t,r}\left(\ddot{\mathbf{s}}_{i}^{\prime}\right)\sum_{k}\left(\frac{\ddot{o}_{r,kt}E_{H_{r,t}\mid\ddot{\mathbf{S}}_{r,t},\lambda}\left(H_{r,t}^{-1}\right)}{\left[\mathbf{b}\widehat{\ddot{\mathbf{s}}_{i}^{\prime}}\right]_{k}}+\hat{\beta}_{k}\ln\left[\mathbf{b}\widehat{\ddot{\mathbf{s}}_{i}^{\prime}}\right]_{k}\right)\nonumber \\
=Q'+\sum_{i}\left(P_{1}\left(\widehat{\ddot{\mathbf{s}}_{i}^{\prime}}\right)+P_{2}\left(\widehat{\ddot{\mathbf{s}}_{i}^{\prime}}\right)\right),
\label{eq:EM_cots_noise}
\end{gather}
where $Q'$ is independent of the state variables $\ddot{\mathbf{s}}^{\prime}$. Due to the summation, it is optimal to minimize \eqref{eq:EM_cots_noise} w.r.t. each $\widehat{\ddot{\mathbf{s}}_{i}^{\prime}}$ independently. It can be easily shown that $P_{1}$ is a convex function where $P_{2}$
is a concave function. Using the CCCP procedure, a convex problem is generated as:
\begin{gather}
\widehat{\ddot{\mathbf{s}}_{i}^{\prime}}\left(l+1\right)=\argmin_{\mathbf{x}}\, P_{1}\left(\mathbf{x}\right)+\mathbf{x}^{\top}\nabla P_{2}\left(\widehat{\ddot{\mathbf{s}}_{i}^{\prime}}\left(l\right)\right),\label{eq:CCCP_problem}\\
s.t.\qquad\qquad\mathbf{x}\geq0\nonumber
\end{gather}
where $\nabla P_{2}\left(\mathbf{s}\right)$ represents the gradient of $P_{2}$ evaluated at $\mathbf{s}$, and $l$ is the iteration number. This constrained problem can be solved by
usual convex optimization tools, e.g. the interior-point methods \cite[ch.11]{Boyd2004}. This procedure is followed iteratively until a locally optimal solution
$\widehat{\ddot{\mathbf{s}}_{i}^{\prime}}$ is obtained. Even though the CCCP procedure does not lead to a closed form solution, it makes
the solution faster and more robust. The required derivatives to solve the above problem are given in Appendix \ref{sec:Gradient-and-Hessian}.
In summary, the following algorithm is pursued to find the optimal babble state value vectors:
\begin{enumerate}
\item Initialize $\widehat{\ddot{\mathbf{s}}_{i}^{\prime}}\left(1\right)$
using the solution obtained in the previous iteration of the EM algorithm
for $i\in\{ 1,\ldots\ddot{N}\} $, set $l=1$.
\item For each $i$, iterate between the following steps until convergence
(usually 2--3 iterations are enough):
\begin{enumerate}
\item Calculate $\nabla P_{2}(\widehat{\ddot{\mathbf{s}}_{i}^{\prime}}\left(l\right))$
(\ref{eq:gradient_CCCP}).
\item Solve problem (\ref{eq:CCCP_problem}) using the interior-point methods.
\item $l=l+1.$
\end{enumerate}
\end{enumerate}
Since the parameter estimation framework is based on the EM, initialization of the algorithm is important. To assign the initial values for $\ddot{\mathbf{s}}_{i}^{\prime}$ (before the first iteration of the EM), we generated two minutes of 10-person babble noise using the TIMIT database
($2\text{f}$+$2\text{m}$ + $1\text{f}{}_{-1.25\text{dB}}$+$1\text{f}{}_{-3\text{dB}}$+$1f{}_{-6\text{dB}}$+$1\text{m}{}_{-1.25\text{dB}}$+$1\text{m}{}_{-3\text{dB}}$+$1\text{m}{}_{-6\text{dB}}$),
and the gamma-HMM (Subsection \ref{sub:Gamma-HMM-as}) was applied independently to each speaker's spectrogram to find the NMF weighting vector $\mathbf{u}{}_{t}$.
Then, all of the ten $\mathbf{u}{}_{t}$ vectors were summed together to obtain $\mathbf{u}{}_{t}^{\text{babble}}$. At the end, a K-means
clustering procedure was applied to cluster all of the columns of $\mathbf{u}^{\text{babble}}$ into $\ddot{N}$ groups whose mean values were used to initialize the state value vectors ($\ddot{\mathbf{s}}_{i}^{\prime}$) for the babble training.
\vspace{-3mm}
\subsection{Updating Time-varying Parameters}
\vspace{-.5mm}
The scale parameters of the stochastic gains are time-variant and, thus, for the purposes of noise reduction they have to be estimated online
given only the noisy signal. In the following, the parameters $\lambda_{t}=\left\{ \theta_{t},\gamma_{t}\right\} $
are updated in a recursive manner after the estimation of the clean speech signal. Therefore, given the noisy signal, a correction term is
calculated and is added to the current estimates to obtain the new estimate of the parameters. In the remainder of this section, this
correction term is obtained and is used to update the time-varying parameters (Eq. (\ref{eq:update_online_theta}) and (\ref{eq:update_online_gamma})).
An algorithm was presented in \cite{Krishnamurthy1993} to estimate the HMM parameters online, that was based on the recursive EM algorithm
and the stochastic approximation \cite{Titterington1984,Weinstein1990}. Here, we follow a similar procedure as described in \cite{Krishnamurthy1993,Zhao2007}
to update $\lambda_{t}$. The recursive EM algorithm is a stochastic approximation in which the parameters of interest are updated sequentially. To do
so, the EM help function is defined as the conditional expectation of the log likelihood of the complete data until the current time
w.r.t. posterior distribution of the hidden variables. Then, this help function is maximized over the parameters by a single-iteration
stochastic approximation in each time instance. Denote the hidden random variables of the EM algorithm as $\mathbf{Z}_{t}=\left\{\mathbf{S}_{t},G_{t},H_{t}\right\}$. Given a noisy observation at time $t$, $\mathbf{y}_t$, a new estimate of the parameters $\hat{\lambda}_{t}$ is obtained by solving:
\begin{gather}
\hat{\lambda}_{t}=\argmax_{\lambda_{t}}\, Q_{t}\left(\lambda_{t},\hat{\lambda}_{1}^{t-1}\right),\qquad\text{where}\nonumber
\end{gather}
\begin{gather}
Q_{t}\left(\lambda_{t},\hat{\lambda}_{1}^{t-1}\right)=\sum_{\mathbf{s}_{1}^{t}}\int\int f\left(\mathbf{z}_{1}^{t}\mid\mathbf{y}_{1}^{t},\hat{\lambda}_{1}^{t-1}\right)\times\nonumber \\
\ln f\left(\mathbf{y}_{1}^{t},\mathbf{z}_{1}^{t}\mid\lambda_{t}\right)d\mathbf{g}_{1}^{t}d\mathbf{h}_{1}^{t},
\label{eq:online_cost}
\end{gather}
where $\hat{\lambda}_{1}^{t-1}=\left\{ \hat{\lambda}_{1},\ldots\hat{\lambda}_{t-1}\right\} $, $\mathbf{z}_{1}^{t}=\left\{ \mathbf{z}_{1},\mathbf{z}_{2},\ldots,\mathbf{z}_{t}\right\} $.
As it is shown in \cite{Krishnamurthy1993}, the objective function
in (\ref{eq:online_cost}) can be simplified, up to an additive constant, as:
\[
Q_{t}\left(\lambda_{t},\hat{\lambda}_{1}^{t-1}\right)=\text{const.}+\sum_{\tau=1}^{t}\mathbb{\mathcal{L}}_{\tau\mid t}\left(\lambda_{t},\hat{\lambda}_{1}^{\tau-1}\right),
\]
with
\begin{gather}
\mathbb{\mathcal{L}}_{\tau\mid t}\left(\lambda_{t},\hat{\lambda}_{1}^{\tau-1}\right)=\sum_{\mathbf{s}_{\tau}}\int\int f\left(\mathbf{z}_{\tau}\mid\mathbf{y}_{1}^{t},\hat{\lambda}_{1}^{\tau-1}\right)\times\nonumber \\
\left(\ln f\left(g_{\tau}\mid\mathbf{s}_{\tau},\lambda_{t}\right)+\ln f\left(h_{\tau}\mid\mathbf{s}_{\tau},\lambda_{t}\right)\right)dg_{\tau}dh_{\tau}.\label{eq:help_func_tau}\end{gather}
The parameters of interest can be updated as \cite{Krishnamurthy1993}:
\begin{equation}
\hat{\lambda}_{t}\hspace{-.15em}=\hspace{-.15em}\hat{\lambda}_{t-1}+\hspace{-.15em}\left.\left(\hspace{-.15em}-\frac{\partial^{2}Q_{t}\left(\lambda_{t},\hat{\lambda}_{1}^{t-1}\right)}{\partial\lambda_{t}^{2}}\hspace{-.15em}\right)^{-1}\frac{\partial\mathbb{\mathcal{L}}_{t\mid t}\left(\lambda_{t},\hat{\lambda}_{1}^{t-1}\right)}{\partial\lambda_{t}}\right|_{\hat{\lambda}_{t-1}}\hspace{-3mm}.
\label{eq:update_online}
\end{equation}
Using the Bayes rule, the posterior probability of the hidden states can be written as:
\begin{gather}
f\left(\mathbf{z}_{\tau}\mid\mathbf{y}_{1}^{t},\hat{\lambda}_{1}^{\tau-1}\right)=\frac{f\left(\mathbf{z}_{\tau},\mathbf{y}_{\tau}\mid\mathbf{y}_{1}^{\tau-1},\mathbf{y}_{\tau+1}^{t},\hat{\lambda}_{1}^{\tau-1}\right)}{f\left(\mathbf{y}_{\tau}\mid\mathbf{y}_{1}^{\tau-1},\mathbf{y}_{\tau+1}^{t},\hat{\lambda}_{1}^{\tau-1}\right)}=\nonumber \\
\frac{f\left(\mathbf{s}_{\tau}\mid\mathbf{y}_{1}^{\tau-1},\mathbf{y}_{\tau+1}^{t},\hat{\lambda}_{1}^{\tau-1}\right)f\left(\mathbf{y}_{\tau},h_{\tau},g_{\tau}\mid\mathbf{s}_{\tau},\hat{\lambda}_{1}^{\tau-1}\right)}{f\left(\mathbf{y}_{\tau}\mid\mathbf{y}_{1}^{\tau-1},\mathbf{y}_{\tau+1}^{t},\hat{\lambda}_{1}^{\tau-1}\right)},
\label{eq:post_online_param}
\end{gather}
where the standard Markov chain property (independency of the observations given the hidden states) is used to get the second line. To reduce
the computation effort, (\ref{eq:post_online_param}) is approximated as (this is also done implicitly in \cite{Zhao2007}):
\begin{gather}
f\left(\mathbf{z}_{\tau}\mid\mathbf{y}_{1}^{t},\hat{\lambda}_{1}^{\tau-1}\right)\approx\nonumber \\
\frac{f\left(\mathbf{s}_{\tau}\mid\mathbf{y}_{1}^{\tau-1},\hat{\lambda}_{1}^{\tau-1}\right)f\left(\mathbf{y}_{\tau},h_{\tau},g_{\tau}\mid\mathbf{s}_{\tau},\hat{\lambda}_{1}^{\tau-1}\right)}{f\left(\mathbf{y}_{\tau}\mid\mathbf{y}_{1}^{\tau-1},\hat{\lambda}_{1}^{\tau-1}\right)}.
\label{eq:approx_post_online}
\end{gather}
For $t=\tau$, the above approximation is exact. Using (\ref{eq:Lapplace_app}),
(\ref{eq:observation_likelihood}), and (\ref{eq:approx_post_online})
in (\ref{eq:help_func_tau}) yields:
\begin{gather*}
\mathbb{\mathcal{L}}_{\tau\mid t}\left(\lambda_{t},\hat{\lambda}_{1}^{\tau-1}\right)\approx\sum_{\mathbf{s}_{\tau}}\omega_{\tau}\left(\mathbf{s}_{\tau},\mathbf{y}_{\tau}\right)\times\\
\left(\ln f\left(g_{\tau}^{\prime}\mid\mathbf{s}_{\tau},\lambda_{t}\right)+\ln f\left(h_{\tau}^{\prime}\mid\mathbf{s}_{\tau},\lambda_{t}\right)\right),
\end{gather*}
where $\omega_{\tau}\left(\mathbf{s}_{\tau},\mathbf{y}_{\tau}\right)=\frac{\mathbf{\zeta}_{\tau}\left(\mathbf{s}_{\tau},\mathbf{y}_{\tau}\right)}{\sum_{\mathbf{s}_{\tau}}\mathbf{\zeta}_{\tau}\left(\mathbf{s}_{\tau},\mathbf{y}_{\tau}\right)}$
is the scaled forward variable, and $\mathbf{\zeta}_{t}\left(\mathbf{s}_{t},\mathbf{y}_{t}\right)=\mathbf{\eta}_{t}\left(\mathbf{s}_{t}\right)f\left(\mathbf{y}_{t},g_{t}^{\prime},h_{t}^{\prime}\mid\mathbf{s}_{t}\right)\frac{2\pi}
{\sqrt{\text{det}(A_{\mathbf{s}_t})}}$ as in Subsection \ref{sub:Clean-Speech-Estimator}. Evaluating (\ref{eq:update_online})
for $\theta_{t}$ and $\gamma_{t}$ yields:

\begin{equation}
\hat{\theta}_{t}=\hat{\theta}_{t-1}+\frac{1}{\mathcal{I}_{t}\left(\hat{\theta}_{t-1}\right)}\sum_{\mathbf{s}_{t}}\omega_{t}\left(\mathbf{s}_{t},\mathbf{y}_{t}\right)\left(\frac{-\phi}{\hat{\theta}_{t-1}}+\frac{g_{t}^{\prime}}{\hat{\theta}_{t-1}^{2}}\right),
\label{eq:update_online_theta}
\end{equation}
\begin{equation}
\mathcal{I}_{t}\left(\hat{\theta}_{t-1}\right)=\sum_{\tau=1}^{t}\sum_{\mathbf{s}_{\tau}}\omega_{\tau}\left(\mathbf{s}_{\tau},\mathbf{y}_{\tau}\right)\left(\frac{-\phi}{\hat{\theta}_{t-1}^{2}}+\frac{2g_{t}^{\prime}}{\hat{\theta}_{t-1}^{3}}\right).
\nonumber
\end{equation}
and
\begin{equation}
\hat{\gamma}_{t}=\hat{\gamma}_{t-1}+\frac{1}{\mathcal{I}_{t}\left(\hat{\gamma}_{t-1}\right)}\sum_{\mathbf{s}_{t}}\omega_{t}\left(\mathbf{s}_{t},\mathbf{y}_{t}\right)\left(\frac{-\psi}{\hat{\gamma}_{t-1}}+\frac{h_{t}^{\prime}}{\hat{\gamma}_{t-1}^{2}}\right),
\label{eq:update_online_gamma}
\end{equation}
\begin{equation}
\mathcal{I}_{t}\left(\hat{\gamma}_{t-1}\right)=\sum_{\tau=1}^{t}\sum_{\mathbf{s}_{\tau}}\omega_{\tau}\left(\mathbf{s}_{\tau},\mathbf{y}_{\tau}\right)\left(\frac{-\psi}{\hat{\gamma}_{t-1}^{2}}+\frac{2h_{t}^{\prime}}{\hat{\gamma}_{t-1}^{3}}\right).
\nonumber
\end{equation}
To ensure the required positivity of the step sizes $1/\mathcal{I}_{t}(\hat{\theta}_{t-1})$ in \eqref{eq:update_online_theta} and $1/\mathcal{I}_{t}(\hat{\gamma}_{t-1})$ in \eqref{eq:update_online_gamma} \cite{Krishnamurthy1993,Weinstein1990},  and to take care of the time-variant parameters, we can modify $\mathcal{I}_{t}(\hat{\lambda}_{t-1})$ slightly by adding a restriction and forgetting factors to reduce the effect of the previous observations as \cite{Krishnamurthy1993,Zhao2007}:
\begin{eqnarray}
\mathcal{I}_{t}\left(\hat{\theta}_{t-1}\right) & = & \xi_{\theta}\mathcal{I}_{t-1}\left(\hat{\theta}_{t-2}\right)+\label{eq:online_gain_update_hes_t}\\
 &  & \max\left(\beta_{\theta},\sum_{\mathbf{s}_{t}}\omega_{t}\left(\mathbf{s}_{t},\mathbf{y}_{t}\right)\left(\frac{-\phi}{\hat{\theta}_{t-1}^{2}}+\frac{2g_{t}^{\prime}}{\hat{\theta}_{t-1}^{3}}\right)\right),\nonumber \end{eqnarray}
 \begin{eqnarray}
\mathcal{I}_{t}\left(\hat{\gamma}_{t-1}\right) & = & \xi_{\gamma}\mathcal{I}_{t-1}\left(\hat{\gamma}_{t-2}\right)+\label{eq:online_gain_update_hes_g}\\
 &  & \max\left(\beta_{\gamma},\sum_{\mathbf{s}_{t}}\omega_{t}\left(\mathbf{s}_{t},\mathbf{y}_{t}\right)\left(\frac{-\psi}{\hat{\gamma}_{t-1}^{2}}+\frac{2h_{t}^{\prime}}{\hat{\gamma}_{t-1}^{3}}\right)\right),\nonumber \end{eqnarray}
with $0<\xi_{\theta},\xi_{\gamma}<1$, and $0<\beta_{\theta},\beta_{\gamma}$.
\vspace{-1mm}
\section{Experiments and Results\label{sec:Experiments-and-Results}}
The capability of the proposed models and the performance of the developed noise reduction algorithm is investigated in various ways. In Subsection
\ref{sub:System-Implementation}, the details of the implementation of the proposed system is explained. In Subsection \ref{sub:7.3 Evaluations},
the developed noise reduction scheme is evaluated and compared to state-of-the-art methods using different objective measures and a subjective
listening test. The performance of the developed system is compared to that of the Bayesian NMF (BNMF) based approach \cite{Mohammadiha2012a}
and the ETSI (European Telecommunications Standards Institute) front end Wiener filtering \cite{ETSI_ES_202050_2007}.

In the BNMF approach, to utilize the temporal correlation of the underlying speech and noise signals, the posterior distributions of the NMF coefficients
at the past time instances were widened and applied as the new prior distribution through the Bayesian framework to obtain an MMSE estimator
for the speech signal \cite{Mohammadiha2012a}. A comparison to the BNMF method has been motivated by the analogy of the proposed babble model
and nonnegative matrix factorization. Also, as it is reported in \cite{Mohammadiha2012a}, the BNMF-based noise reduction approach outperforms
different competing algorithms. On the other hand, the ETSI two-stage Wiener filter is
carefully tuned for good performance in denoising speech \cite{ETSI_ES_202050_2007}, and it is considered here to compare the performance of the model-based
approaches to a standard approach that does not benefit from trained noise-specific models.
\vspace{-4mm}
\subsection{System Implementation\label{sub:System-Implementation} }
\vspace{-1mm}
The proposed models for speech and babble signals were trained using the TIMIT and NOISEX-92 databases, respectively. All of the signals were
down-sampled to 16-kHz and the DFT was implemented using a frame length of $320$ samples with $50\%$ overlapped windows using a Hann window.
For speech, $600$ sentences from the training set of the TIMIT were used as training data, and for babble noise the first 75\% of
the signal was used for training while the rest of the signal was used for the test purposes. To investigate the performance of the
algorithms as a function of the number of speakers in the babble noise, a different set of babble training and testing data was used,
which is explained in Subsection \ref{sub:effect_num_speakers_babble}. Also, the core test set of the TIMIT database ($192$ sentences) was
exploited for the noise reduction evaluation. The signal synthesis was performed using the overlap-and-add procedure.

For the speech model, $\bar{N}=55$ states were trained in order to roughly identify these states by different phonemes. For the babble model, a discrete
HMM with $\ddot{N}$ states was considered. As a result, the final mixed signal model includes $N=\bar{N}\ddot{N}$ states. To carry
out the speech enhancement and calculate the final speech gain $\kappa_{kt}$ (\ref{eq:final_gain_se}), the weighted sum of the state-conditional
Wiener filters has to be calculated while for each of the $N$ states, the MAP estimate of the stochastic gain variables has to be performed, which is time consuming in general. Although a large value for $\ddot{N}$ may approximate the underlying continuous state-space of the babble
model better, it will result in a computationally more expensive system and for $\ddot{N}$ larger than $50$ a pruning algorithm \cite{Sameti1998,Zhao2007}
has to be implemented to keep the level of complexity practical. In our experiments, we set $\ddot{N}=10$ (except Subsection \ref{sub:effect_num_speakers_babble}) since the performance was quite similar for $\ddot{N}$ in the range of 10 to 200.
Moreover, we observed that a high shape parameter (5$\sim$30) for the stochastic gain variables makes the MAP estimation faster while the performance remains similar. Hence, in our simulations the shape parameters of both the stochastic gain variables were set to 15 although the data-driven estimate of the shape parameter of the speech stochastic gain variable was less than one (this is an indication of a high variation in the state-conditioned energy level of the signal). For this setup, our Matlab implementation runs in approximately 5-times real time\footnote{By real time, we mean that the processing of the current frame finishes before the next frame arrives.} using a PC with 3.8 GHz Intel CPU and 2 GB RAM. The online parameter estimation (\ref{eq:online_gain_update_hes_t},\ref{eq:online_gain_update_hes_g})
was done using $\xi_{\theta}=0.99$, $\xi_{\gamma}=0.98$, and $\beta_{\theta}=\beta_{\gamma}=100$, which were set experimentally.

Additionally, motivated by our previous work \cite{Mohammadiha2011a}, an exponential smoothing was performed as $\bar{\kappa}_{kt}=0.4\bar{\kappa}_{k\left(t-1\right)}+0.6\kappa_{kt}$ and the speech signal was estimated as $\hat{x}_{kt}=\bar{\kappa}_{kt}y_{kt}$. This smoothing slightly improves the quality of the estimated speech
signal by smoothing out the gain fluctuations. For the BNMF approach \cite{Mohammadiha2012a}, $60$ basis vectors for speech and $100$ basis
vectors for babble were trained using the same training material as explained above. For this method, an informative prior
was only used for babble NMF coefficients since applying informative prior for speech NMF coefficients did not result in better noise reduction
performance, as also mentioned in \cite{Mohammadiha2012a}.
\vspace{-3mm}
\subsection{Evaluations\label{sub:7.3 Evaluations}}
\vspace{-1mm}
In this section, we evaluate the proposed system and compare its performance with that of BNMF \cite{Mohammadiha2012a} and ETSI front end Wiener
filtering \cite{ETSI_ES_202050_2007}. First we present a general comparison of methods, and then some specific aspects are highlighted.
Finally, the results of the subjective listening tests are given.
\subsubsection{Objective Evaluation of the Noise Reduction\label{sub:Objective-Evaluation}}
Five different objective measures were considered for the evaluation: (1) source to distortion ratio\textit{ (SDR)} \cite{Vincent2006}
that represents the overall quality of speech; (2) long-term signal to noise ratio (\textit{SNR}); (3) segmental SNR (\textit{SegSNR})
\cite[ch. 10]{Loizou2007}, which was limited to the range {[}$-10$ dB, $30$ dB{]}; (4) spectral distortion (\textit{SD}) \cite{Paliwal1995},
for which the time-frames with powers $40$ dB less than the long-term power level were excluded from the evaluations; (5) perceptual evaluation
of speech quality (\textit{PESQ}) \cite{PESQ2000}. The evaluation is performed at three input \textit{SNR}s: 0, 5, and 10 dB.

\begin{figure}
\centering
\includegraphics[width=9cm]{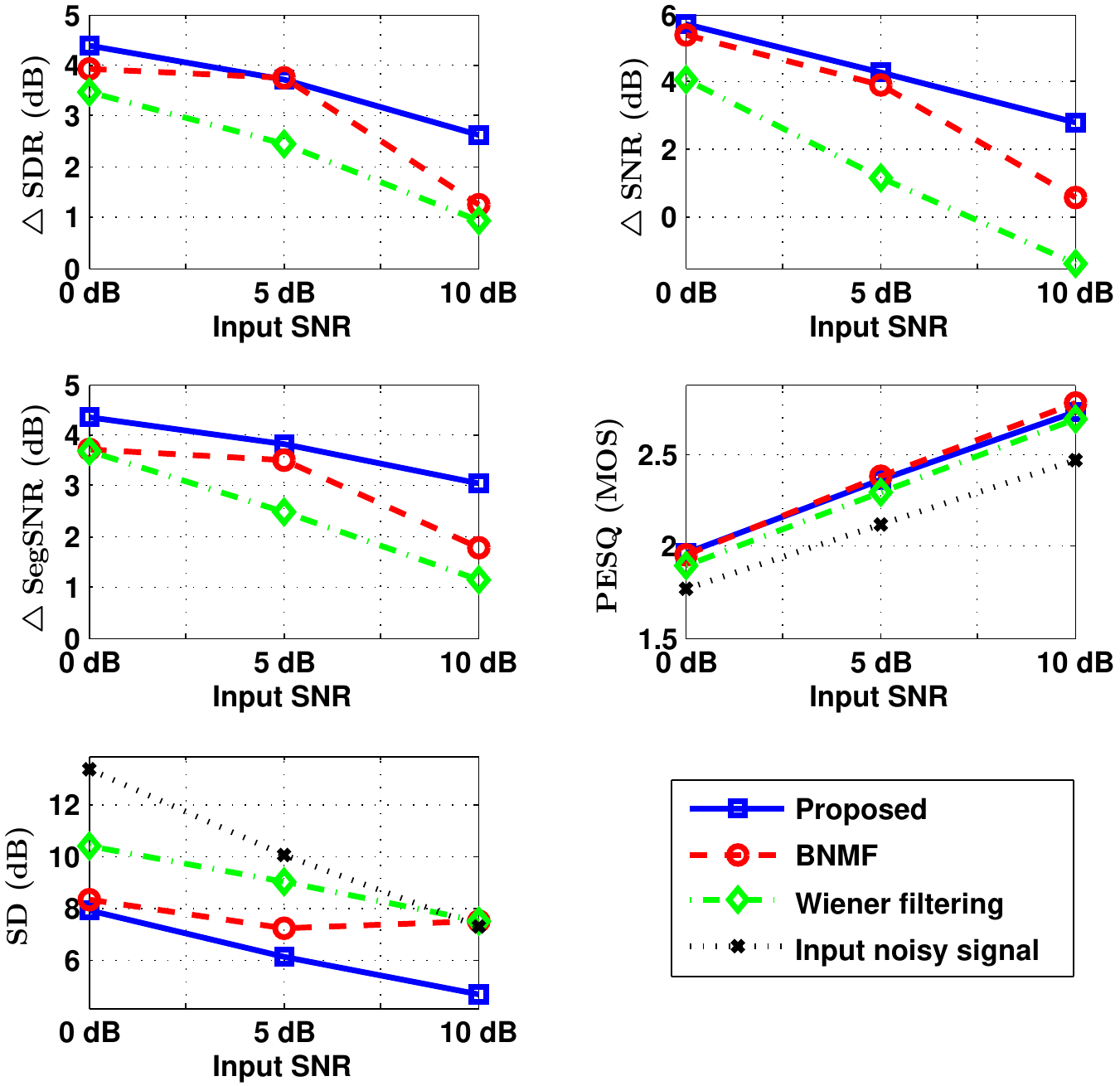}
\vspace{-4mm}
\caption{Objective evaluation of the application of the proposed babble model
in noise reduction. $\triangle$ is used to show the improvements gained by the noise reduction algorithms, e.g., $\triangle \textit{SDR}=\textit{SDR}_{\text{enhanced}}-\textit{SDR}_{\text{noisy}}$.}%
\label{fig:obj_results}
\vspace{-2mm}
\end{figure}

The results are presented in Fig.~\ref{fig:obj_results}. For \textit{SDR}, \textit{SNR}, and \textit{SegSNR} the improvements in dB (e.g. $\triangle \textit{SDR}=\textit{SDR}_{\text{enhanced}}-\textit{SDR}_{\text{noisy}}$) are shown in this figure for readability.  For \textit{PESQ} and \textit{SD} the actual values for the enhanced signals and for the noisy input signal are shown. A high degree of consistency can be seen between the different measures. The results show that the two model-based approaches lead to much better improvements than the Wiener filtering. The proposed method outperforms the BNMF in all of the input \textit{SNR}s in the sense of
\textit{SDR}, \textit{SNR}, \textit{SegSNR}, and \textit{SD}. For \textit{PESQ}, gamma-NHMM results to slightly better \textit{PESQ} improvement at 0 dB input \textit{SNR}, while BNMF leads to slightly better improvements at 5 and 10 dB \textit{SNR}s. However, the difference between \textit{PESQ} values for two algorithms is marginal in all three \textit{SNR}s.
\subsubsection{Effect of Systems on Speech and Noise Separately\label{sub:ShadowFiltering}}
A desired feature of a noise reduction system is that the speech signal remains undistorted. In order to compare this aspect of the algorithms,
segmental speech \textit{SNR} ($\textit{SNR}_{\text{seg-sp}}$), and\textit{ }segmental noise reduction (\textit{SegNR}) \cite{Lotter2005} were measured in a shadow
filtering framework. Hence, the enhancement filter was obtained using the input noisy signal (as it was done in \ref{sub:Objective-Evaluation}),
and it was applied to the clean speech and noise components of the input noisy signal, separately. The output speech and noise signals
were compared to the corresponding inputs to compute these two measures. For both measures a high value is desired, and $\textit{SNR}_{\text{seg-sp}}$ is
inversely proportional to the speech distortion.

\begin{figure}
\centering
\includegraphics[scale=0.55]{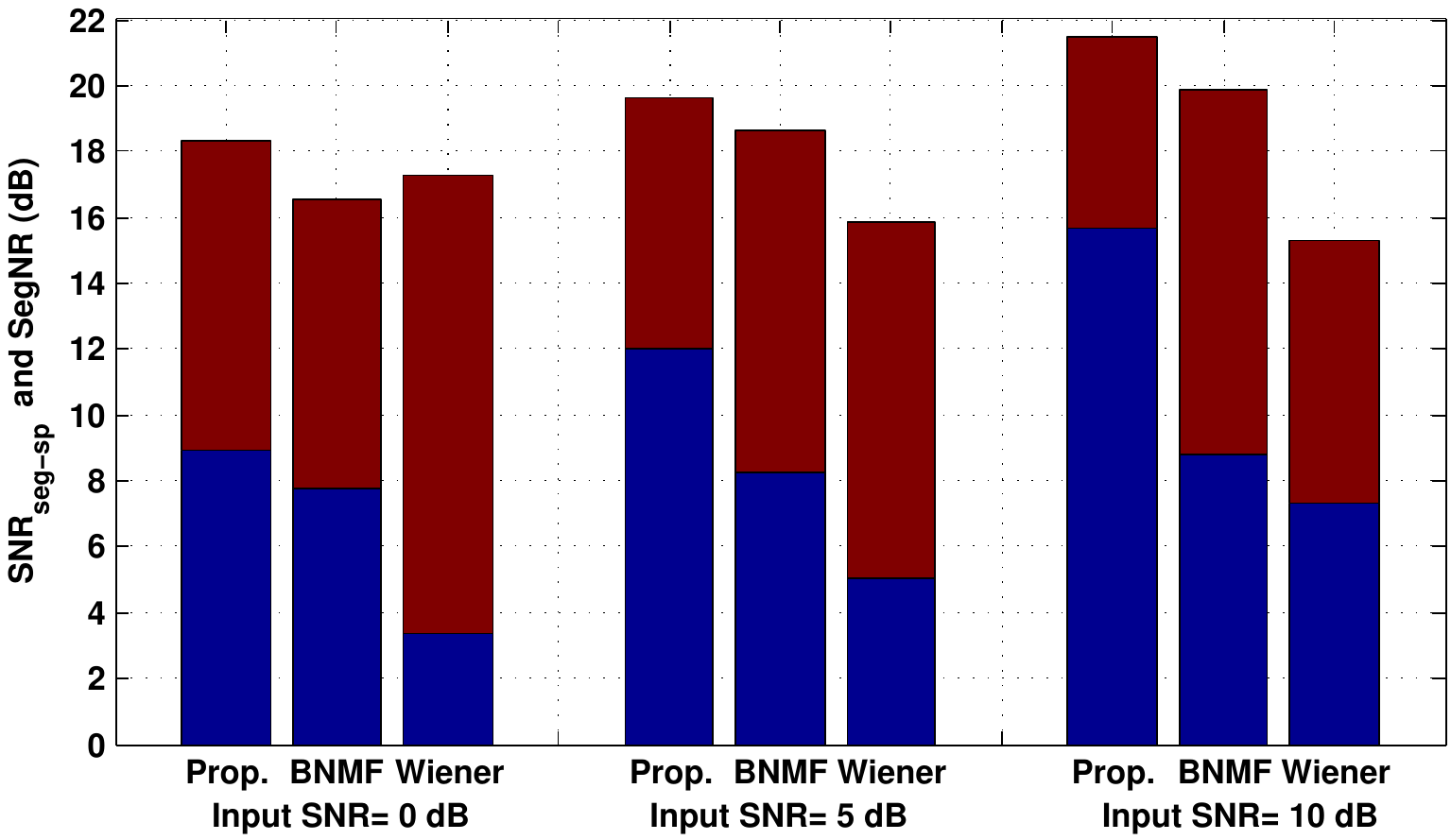}\caption{Stacked presentation of the segmental speech \textit{SNR} ($\textit{SNR}_{\text{seg-sp}}$, bottom),
and the segmental noise reduction (\textit{SegNR}, top). \label{fig:shadow_filt_results}}
\vspace{-5mm}
\end{figure}

The results are shown in Fig.~\ref{fig:shadow_filt_results}. As it can be seen in the figure,
the proposed system leads to a higher segmental speech \textit{SNR} (less distortion) in all of the input \textit{SNR}s. Also, the sum of
the $\textit{SNR}_{\text{seg-sp}}$ and \textit{SegNR} is the highest for the proposed method.
\subsubsection{Effect of the Number of Speakers in Babble\label{sub:effect_num_speakers_babble}}
It is well known that the performance of the model-based noise reduction systems that are trained for a specific signal degrades when there
is a mismatch between training and testing. Therefore, in the case of a mismatch, the
standard Wiener filter might outperform the model-based approaches since it is not restricted to any specific noise type. In this part,
we investigate the performance of the noise reduction algorithms as a function of the number of speakers in the babble. For the experiments,
an artificial babble was generated by adding waveforms of different speakers from the TIMIT database, with equal speech level for all of the
speakers. The number of speakers in generating babble were chosen as $M\in\{4,6,10,20,50,100\}$. Moreover, the babble noise from NOISEX-92 is also considered in the evaluation for comparison.

For the proposed method and BNMF, two babble models were trained using (1) only the NOISEX-92, (2) both the NOISEX-92 and 10-person babble noise
(different from the test signal). Also, for the proposed system, we trained $\ddot{N} = 50$ states for the babble for both of the models
since a pilot experiment indicated that $\ddot{N} = 10$ was insufficient in this case
because of the high non-stationarity of the noise.

\begin{figure}
\centering
\subfloat[\label{fig:trained_for_noisex} Babble models trained using only NOISEX-92. The results are shown for two input \textit{SNR}s 5 dB (left) and
10 dB (right).]{\includegraphics[scale=0.57]{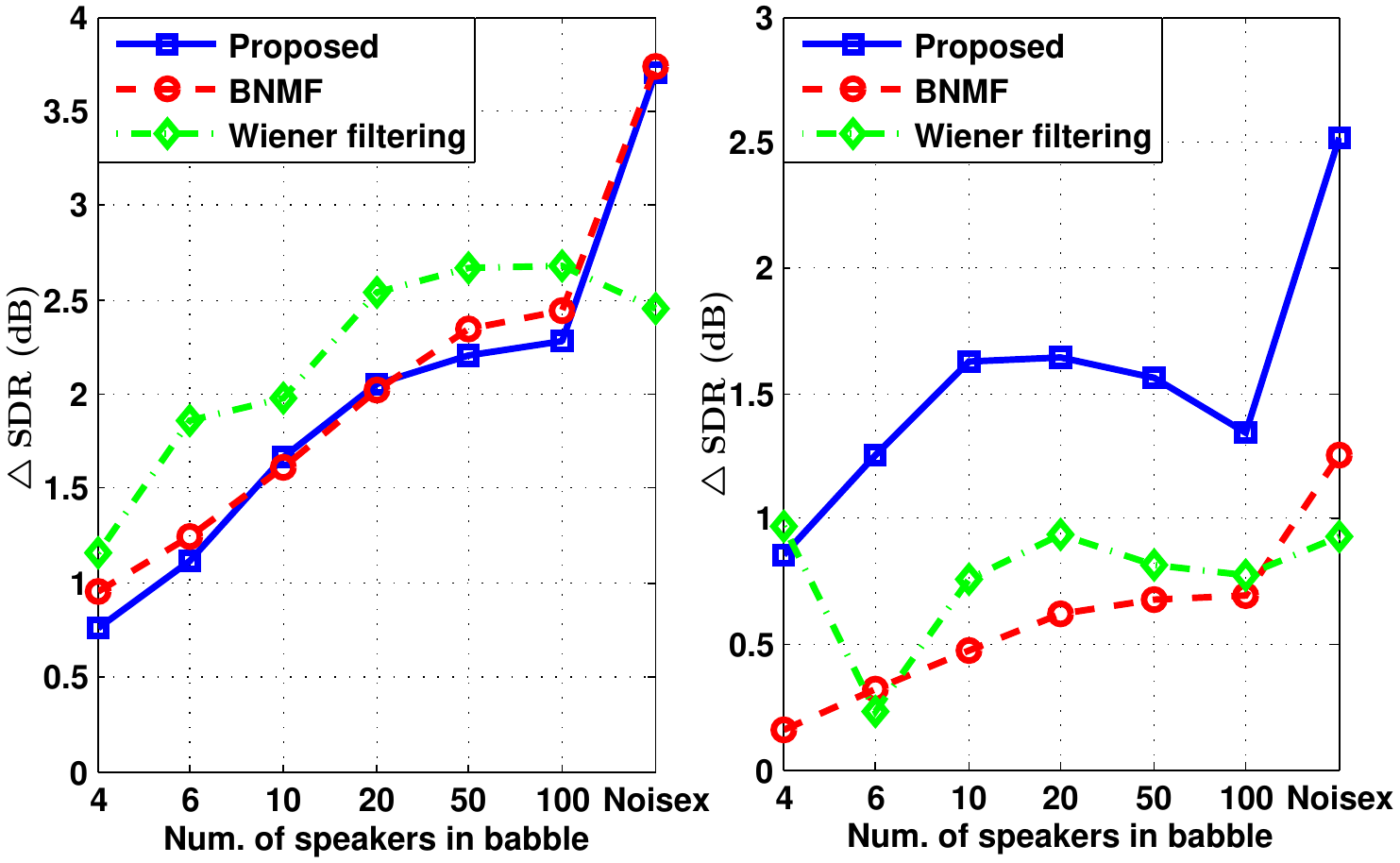}
}

\subfloat[\label{fig:trained_jointly}Babble models trained using both the NOISEX-92 and 10-person babble (different from the testing signals). The results
are shown for two input \textit{SNR}s 5 dB (left) and 10 dB (right).]{\includegraphics[scale=0.57]{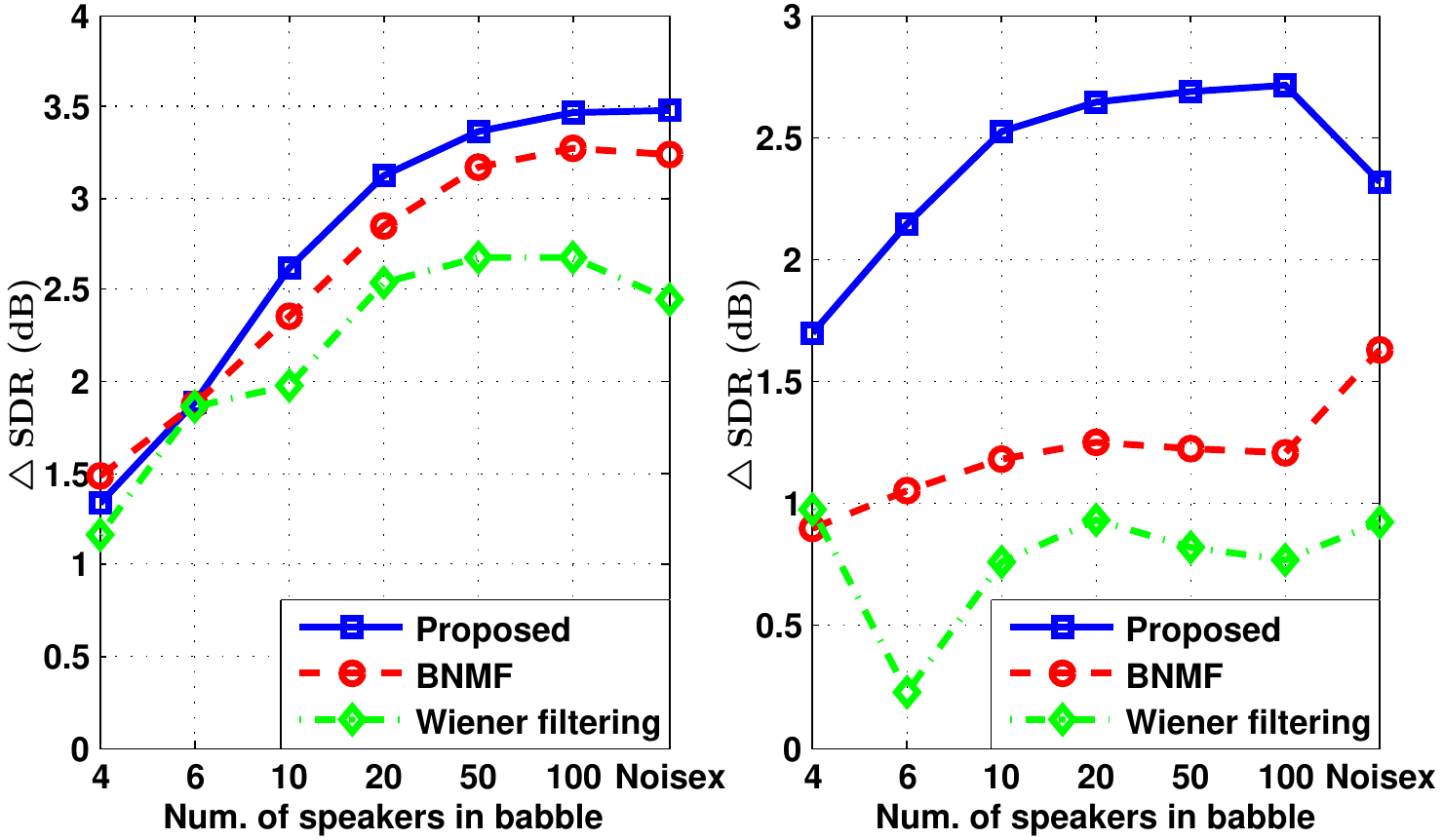}
}\caption{\label{fig:different_number_speakers}Performance of the noise reduction algorithms as a function of the number of speakers in the babble.}
\vspace{-3mm}
\end{figure}

Improvements gained in the source to distortion ratio\textit{ ($\triangle$SDR)} are shown in Fig.~\ref{fig:different_number_speakers} for two input
\textit{SNR}s, 5 and 10 dB. Fig.~\ref{fig:trained_for_noisex} shows the results using the babble model that is trained on only the NOISEX-92.
Looking at the 5 dB input \textit{SNR} scenario (left-hand side of Fig.~\ref{fig:trained_for_noisex}), it can be seen that even though
the performance of the model-based approaches is much better when exposed to the NOISEX-92 babble noise, the ETSI Wiener filter gives
a better result in the other types of babble noise. Fig.~\ref{fig:trained_jointly} shows the results using the babble model that is trained using both the
NOISEX-92 and 10-person artificial babble noise. Here, the performance of the model-based approaches is slightly reduced for the NOISEX-92
babble noise, but in general their performance is significantly improved (especially for the proposed method). This also implies that if the proposed
method is combined with another system that estimates the number of the speakers from the observed babble signal (for example \cite{Krishnamurthy2009}), the performance might be improved further.

\subsubsection{Cross-predictive Test for Model Fitting\label{sub:cross-predictive-test}}
A cross-predictive test was carried out in which both of the speech and babble models from the proposed and the BNMF frameworks
were applied to the speech and babble signals (as separate inputs) in a predictive way. Here, the goal is to investigate whether the
babble (speech) model fits to the babble (speech) signal better than the speech (babble) model. Two measures were computed to compare the
input and the estimated signals. To compare the signals in the spectral domain, spectral distortion (\textit{SD}) \cite{Paliwal1995} was
measured. To compare the input and estimated waveforms, segmental SNR (\textit{SegSNR}) \cite[ch. 10]{Loizou2007} was measured. To
reconstruct the output waveforms, the NMF representations of the spectrograms together with the phase information from the input signal were fed
into the inverse DFT. For the proposed method, the gamma-NHMM representation (similar to Subsection \ref{sub:Gamma-HMM-as}) was used to obtain
the NMF approximation, and for the BNMF that was achieved by multiplying the mean values of the posterior distributions of the basis matrix
and the coefficients matrix.
\begin{table}
\centering
\caption{\label{tab:Confusion_matrices}A cross-predictive test for the different models. The specified signal is fed as the input signal to the given
model and the quality of the reconstructed signal is measured. The results are averaged over the test set explained in Subsection \ref{sub:System-Implementation}.}
\subfloat[Spectral distortion (\textit{SD}, lower value is desired) in dB.]{%
\begin{tabular}{|c|c|c|}
\hline
\multirow{2}{*}{Proposed} & Speech  & Babble \tabularnewline
 & Model & Model\tabularnewline
\hline
Speech Sig. & 3.9 & 6.7\tabularnewline
\hline
Babble Sig. & 3.2 & 2.4\tabularnewline
\hline
\end{tabular} %
\begin{tabular}{|c|c|c|}
\hline
\multirow{2}{*}{BNMF} & Speech  & Babble\tabularnewline
 & Model & Model\tabularnewline
\hline
Speech Sig. & 6.3 & 9.4\tabularnewline
\hline
Babble Sig. & \color{red}{2.2} & \color{red}{2.8}\tabularnewline
\hline
\end{tabular}
}

\subfloat[Segmental SNR (\textit{SegSNR}, higher value is desired) in dB.]{%
\begin{tabular}{|c|c|c|}
\hline
\multirow{2}{*}{Proposed} & Speech  & Babble \tabularnewline
 & Model & Model\tabularnewline
\hline
Speech Sig. & 3.2 & -2.1\tabularnewline
\hline
Babble Sig. & 4 & 5.3\tabularnewline
\hline
\end{tabular} %
\begin{tabular}{|c|c|c|}
\hline
\multirow{2}{*}{BNMF} & Speech  & Babble \tabularnewline
 & Model & Model\tabularnewline
\hline
Speech Sig. & 9.1 & -2.2\tabularnewline
\hline
Babble Sig. & \color{red}{9.2} & \color{red}{6.5}\tabularnewline
\hline
\end{tabular}
}
\vspace{-6mm}
\end{table}

The results of this predictive test are shown in TABLE~\ref{tab:Confusion_matrices} in the form of confusion matrices. If a model is good then for each type of signal (each row in the table),
the best result should be found in the element on the main diagonal.
Both of the measures point in the same
direction, and show that in the proposed framework a better score is obtained for the speech and babble signals using the speech
and babble models, respectively.
However, for the BNMF, the speech model gives a better score to the babble signal than the babble model
itself (shown in a red color in the table). This is because the babble spectrogram can be approximated quite well by combining the speech
basis vectors freely. The result of this test is another indication of the excellence of the proposed babble model, and provides an additional explanation for the achieved results in the previous subsections.
\subsubsection{Subjective Evaluation of the Noise Reduction}
\begin{figure}[t]
\centering
\includegraphics[scale=0.55]{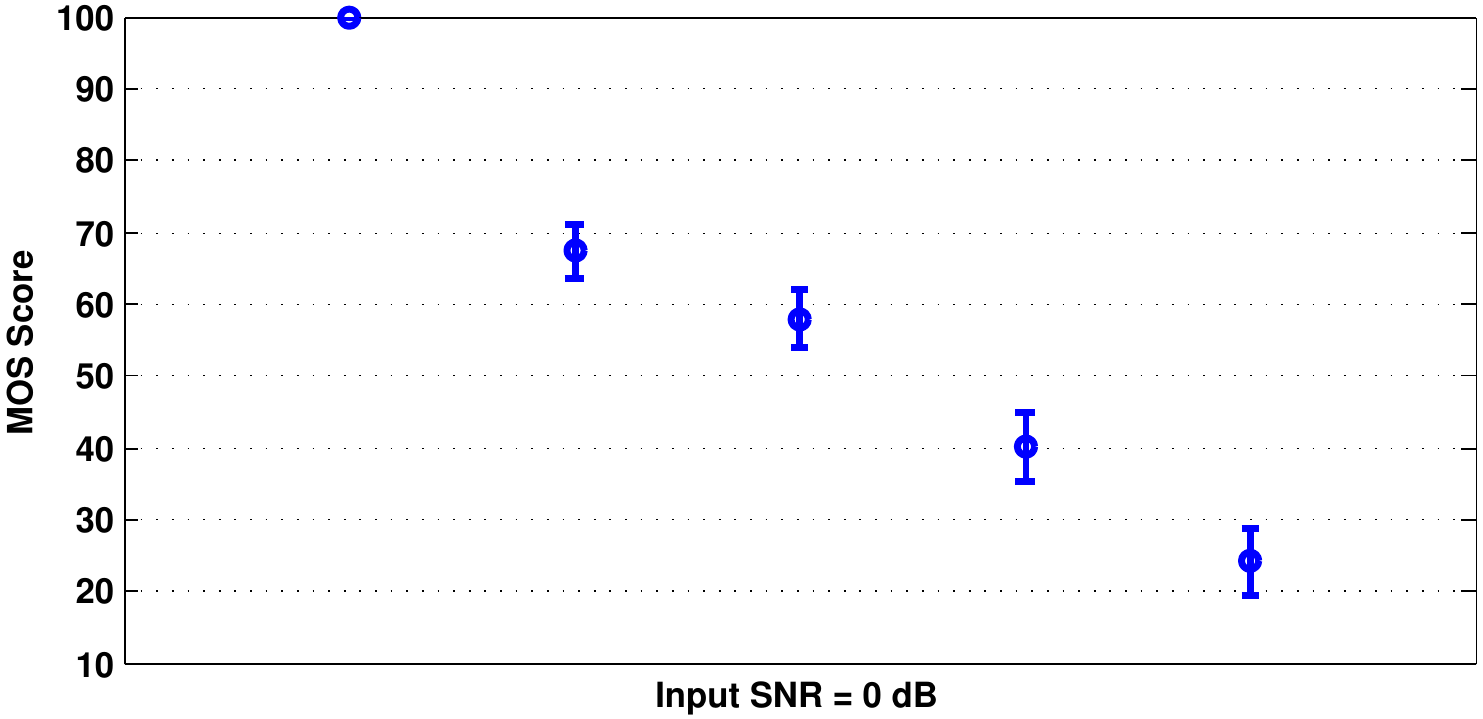}
\includegraphics[scale=0.55]{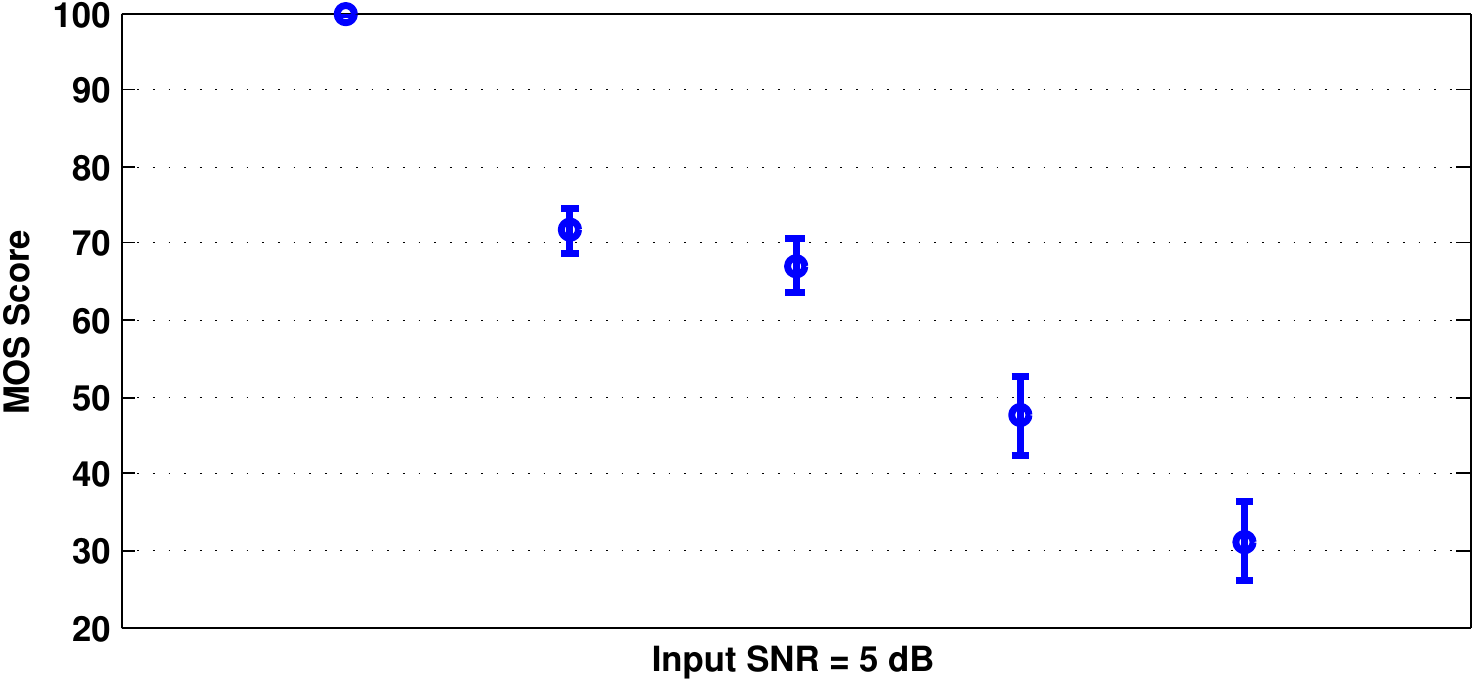}
\includegraphics[scale=0.55]{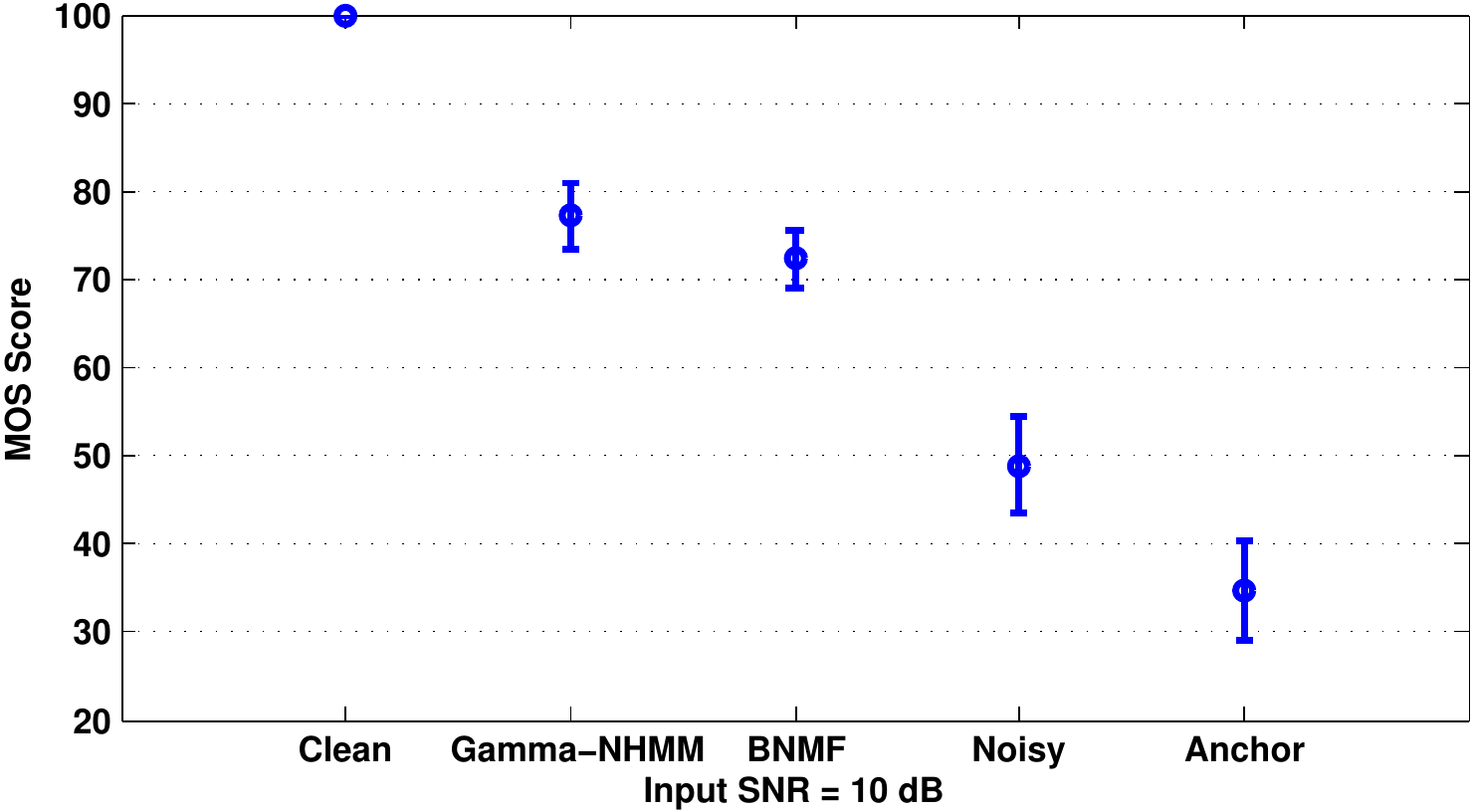}
\vspace{-1mm}
\caption{\label{fig:Results-of-MUSRA}Results of the MUSHRA test at 3 input \textit{SNR}s: 0 dB, 5 dB, 10 dB (top to down) with a 95\% confidence interval.}%
\vspace{-7mm}
\end{figure}
To assess the subjective quality of the estimated speech signal, a subjective listening test was carried out. The test setup was similar to the
ITU recommendation ITU-R BS.1534-1 MUSHRA \cite{Mushra_test}. Six experienced and four inexperienced listeners (ten in total) participated
in the test. The subjective evaluation was performed for three input \textit{SNR}s (0, 5, 10 dB), and for each \textit{SNR} seven sentences
from the core test set of the TIMIT database (4 males and 3 females) were presented to the listeners. In each of the 21 listening sessions,
5 signals were compared by the listeners: (1) reference clean speech signal, (2) noisy speech signal, (3,4) estimated speech signals using
the gamma-NHMM and BNMF, and (5) a hidden anchor signal that was chosen to be the noisy signal at a 5 dB lower \textit{SNR} than the noisy signal
processed by the systems (as suggested in \cite{Volodya2006}). The listeners were allowed to play each sentence as many times as they
wanted, and they always had access to the reference signal. They were asked to rate the signals based on the global speech quality.
Also, some sample signals were presented, and the graphical user interface was introduced to the listeners prior to the test procedure. The order of the signals was randomized with respect to the algorithm
and input \textit{SNR}. Each listener took around 30 minutes on average to complete the listening test.

The results of this listening test, averaged over all of the participants, with a 95\% confidence interval are presented in Fig.~\ref{fig:Results-of-MUSRA}.
At all of the three \textit{SNR}s the gamma-NHMM was preferred over the BNMF algorithm. For 0 dB, the difference is $9.5$ MOS units, whereas for 5 dB and 10 dB, the preference is around $5$ on a scale of $100$. Also, both of the algorithms were preferred over the noisy input signal
by at least $20$ units. According to the spontaneous comments by the listeners, the remaining noise and artifacts in the enhanced signal by the gamma-NHMM is more like
a natural babble-noise while the artifacts introduced by the BNMF are more artificially modulated.

To verify the statistical significance
of the preference of the gamma-NHMM algorithm, a one-tailed t-test was performed. This statistical analysis shows that the gamma-NHMM
leads to a significantly better performance than the BNMF at all three \textit{SNR}s. For 0 dB, the significance level was $p\approx9.10^{-5}$, for 5 dB it was $p\approx0.013$, and finally for 10 dB we obtained $p\approx0.014$.
\vspace{-2mm}
\section{Conclusion\label{sec:Conclusion}}
\vspace{-.5mm}
As babble noise is generated by adding some different speech signals, improving the intelligibility and quality of the speech signal degraded by the babble noise has been a challenging task for a long time. In this paper, a gamma nonnegative HMM was proposed to model the normalized power spectra of babble noise in which the babble basis vectors were
identical to the speech basis vectors. In the proposed models, the time-varying energy levels of speech and babble signals were modeled by gamma distributions whose scale parameters were estimated online.

 The simulations show that the proposed system achieves better model recognition (i.e. babble signal gets a better score with the babble model rather than the speech model) compared to the Bayesian NMF approach. Also, the objective evaluations and the subjective MUSHRA listening test verify the excellence of the proposed noise reduction system. For instance, at $0$ dB input \textit{SNR}, the enhanced speech of the currently developed system was preferred by around $10$ MOS units to the enhanced speech of the closest competing algorithm (Bayesian NMF) and by $27$ to the input noisy signal in the scale of $100$. Moreover, the simulations show that the proposed noise reduction scheme is less sensitive to a mismatch (varying number of speakers in babble) compared to the other competing model-based approach.
\appendices
\vspace{-2mm}
\section{MAP Estimate of the Gain Variables\label{sec:MAP-Estimate-of}}
\vspace{-1mm}
Problem (\ref{eq:max_for_gains}) is a MAP estimator that can be solved by the standard EM algorithm. Let the hidden variables for
EM be $\mathbf{Z}=\left\{ \mathbf{X}_{t},\mathbf{V}_{t}\right\}$, and $\lambda=\left\{ g_{t}^{\prime},h_{t}^{\prime}\right\} $ be the
parameters of interest. Thus, the EM help function is written to $Q(\hat{\lambda},\lambda)=E_{\mathbf{Z}\mid\mathbf{Y}_{t},\lambda}(\ln f(\mathbf{y}_{t},\mathbf{x}_{t},\mathbf{v}_{t}\mid\mathbf{s}_{t},\hat{\lambda})+\ln f(\hat{\lambda}))$.
The terms containing $\hat{g_{t}^{\prime}}$ can be gathered into
\begin{equation}
Q_{\hat{g_{t}^{\prime}}}\left(\hat{\lambda},\lambda\right)=E_{\mathbf{Z}\mid\mathbf{Y}_{t},\lambda}\left(\ln f\left(\mathbf{x}_{t}\mid\hat{g_{t}^{\prime}},\bar{S}_{t}=i\right)+\ln f\left(\hat{g_{t}^{\prime}}\right)\right).
\label{eq:cost_MAP_g}
\end{equation}
Taking the derivative of (\ref{eq:cost_MAP_g}) and setting it to zero yields the solution:
\begin{equation}
\hat{g_{t}^{\prime}}\hspace{-.15em}=\hspace{-.15em}\frac{-\left(K\theta_{t}-\theta_{t}\left(\phi-1\right)\right)\hspace{-.15em}+\hspace{-.15em}\sqrt{\hspace{-.1em}\left(K\theta_{t}-\theta_{t}\left(\phi-1\right)\right)^{2}\hspace{-.1em}+\hspace{-.1em}4\theta_{t}C_{X}}}{2},
\label{eq:MAP_g}
\end{equation}
where $K$ is the number of frequency bins, dimension of $\mathbf{y}_{t}$,
and $C_{X}=\sum_{k=1}^{K}\frac{E\left(\left|X_{kt}\right|^{2}\mid\mathbf{Y}_{t},\lambda\right)}{\alpha_{k}b_{ki}}$.
The posterior expected value of $\left|X_{kt}\right|^{2}$ is calculated using (\ref{eq:wiener_state},\ref{eq:wiener_covariance}). The update
rule for $\hat{h_{t}^{\prime}}$ is also given as:
\begin{equation}
\hat{h_{t}^{\prime}}\hspace{-.15em}=\hspace{-.15em}\frac{-\left(K\gamma_{t}-\gamma_{t}\left(\psi-1\right)\right)\hspace{-.15em}+\hspace{-.15em}\sqrt{\hspace{-.1em}\left(K\gamma_{t}-\gamma_{t}\left(\psi-1\right)\right)^{2}\hspace{-.1em}+\hspace{-.1em}4\gamma_{t}C_{V}}}{2},
\label{eq:MAP_h}
\end{equation}
where $C_{V}=\sum_{k=1}^{K}\frac{E\left(\left|V_{kt}\right|^{2}\mid\mathbf{Y}_{t},\lambda\right)}{\beta_{k}[\mathbf{b}\ddot{\mathbf{s}}_{t}]_k}$.
In very rare cases in practice, the above algorithm may get stuck at a non-maximum
stationary point in which the EM algorithm has to be repeated from a different initial point to obtain a local maximum of the likelihood.

The negative Hessian matrix in (\ref{eq:Lapplace_app}) is defined as:
\begin{gather*}
A_{\mathbf{s}_{t}}(1,1)=-\frac{\partial^{2}\left(\ln f\left(\mathbf{y}_{t},g_{t},h_{t}\mid\mathbf{\mathbf{s}_{t}}\right)\right)}{\partial g_{t}\partial g_{t}}=\frac{\left(\phi-1\right)}{g_{t}^{2}}-\sum_{k=1}^{K}\\
\frac{\left(g_{t}\alpha_{k}\mathbf{b}_{ki}\right)^{2}}{\left(g_{t}\alpha_{k}\mathbf{b}_{ki}+h_{t}\beta_{k}\left[\mathbf{b}\ddot{\mathbf{s}}_{i}^{\prime}\right]_{k}\right)^{2}}\hspace{-.15em}\left(\hspace{-.15em}1-\frac{2\left|y_{kt}\right|^{2}}{\left(g_{t}\alpha_{k}\mathbf{b}_{ki}+h_{t}\beta_{k}\left[\mathbf{b}\ddot{\mathbf{s}}_{i}^{\prime}\right]_{k}\right)}\right)\hspace{-.25em},\\
\end{gather*}
\vspace{-5.5mm}
\begin{gather*}
A_{\mathbf{s}_{t}}(1,2)=A_{\mathbf{s}_{t}}(2,1)=-\frac{\partial^{2}\left(\ln f\left(\mathbf{y}_{t},g_{t},h_{t}\mid \mathbf{\mathbf{s}_{t}}\right)\right)}{\partial g_{t}\partial h_{t}}=-\sum_{k=1}^{K}\\
\frac{\left(g_{t}\alpha_{k}\mathbf{b}_{ki}\right)\left(h_{t}\beta_{k}\left[\mathbf{b}\ddot{\mathbf{s}}_{i}^{\prime}\right]_{k}\right)}{\left(g_{t}\alpha_{k}\mathbf{b}_{ki}+h_{t}\beta_{k}\left[\mathbf{b}\ddot{\mathbf{s}}_{i}^{\prime}\right]_{k}\right)^{2}}\hspace{-.15em}\left(\hspace{-.15em}1-\frac{2\left|y_{kt}\right|^{2}}{\left(g_{t}\alpha_{k}\mathbf{b}_{ki}+h_{t}\beta_{k}\left[\mathbf{b}\ddot{\mathbf{s}}_{i}^{\prime}\right]_{k}\right)}\right)\hspace{-.25em},\\
\end{gather*}
\vspace{-5.5mm}
\begin{gather*}
A_{\mathbf{s}_{t}}(2,2)=-\frac{\partial^{2}\left(\ln f\left(\mathbf{y}_{t},g_{t},h_{t}\mid \mathbf{\mathbf{s}_{t}}\right)\right)}{\partial h_{t}\partial h_{t}}=\frac{\left(\psi-1\right)}{h_{t}^{2}}-\sum_{k=1}^{K}\\
\frac{\left(h_{t}\beta_{k}\left[\mathbf{b}\ddot{\mathbf{s}}_{i}^{\prime}\right]_{k}\right)^{2}}{\left(g_{t}\alpha_{k}\mathbf{b}_{ki}+h_{t}\beta_{k}\left[\mathbf{b}\ddot{\mathbf{s}}_{i}^{\prime}\right]_{k}\right)^{2}}\hspace{-.15em}\left(\hspace{-.15em}1-\frac{2\left|y_{kt}\right|^{2}}{\left(g_{t}\alpha_{k}\mathbf{b}_{ki}+h_{t}\beta_{k}\left[\mathbf{b}\ddot{\mathbf{s}}_{i}^{\prime}\right]_{k}\right)}\right)\hspace{-.25em}.
\end{gather*}
\vspace{-3mm}
\section{Posterior Distribution of the Gain Variables\label{app.Posteior_dist_gain}}
\vspace{-.5mm}
The posterior distribution of the stochastic gain variable of the speech signal in Subsection \ref{sub:Speech-Model-Training}, given the hidden Markov state and the observation is given
by:
\[
f\hspace{-.2em}\left(g_{r,t}\hspace{-.1em}\mid\hspace{-.1em}\bar{\mathbf{o}}_{r,t},\bar{S}_{r,t}=i,\lambda\right)\hspace{-.1em}=\hspace{-.1em}\frac{f\hspace{-.2em}\left(\bar{\mathbf{o}}_{r,t}\hspace{-.1em}\mid \hspace{-.1em} g_{r,t},\bar{S}_{r,t}=i,\lambda\right)\hspace{-.2em}f\hspace{-.2em}\left(g_{r,t}\hspace{-.1em}\mid\hspace{-.1em}\lambda\right)}{f\hspace{-.2em}\left(\bar{\mathbf{o}}_{r,t}\mid\bar{S}_{r,t}=i,\lambda\right)},\]
where $\lambda=\left\{\bar{\mathbf{a}}, \mathbf{b},\boldsymbol{\alpha},\phi,\theta\right\}$ is the estimated parameters from the previous iteration of the Baum-Welch
algorithm. Since the denominator is constant, using (\ref{eq:vector_speech_dist}) and (\ref{eq:gain_speech}) we get:
\begin{gather}
\ln f\left(g_{r,t}\mid\bar{\mathbf{o}}_{r,t},\bar{S}_{r,t}=i,\lambda\right)\propto\nonumber \\
\sum_{k=1}^{K}\left(-\alpha_{k}\ln g_{r,t}-\frac{\bar{o}_{r,kt}}{g_{r,t}b_{k,i}}\right)+\nonumber \\
\left(\phi-1\right)\ln g_{r,t}-\frac{g_{r,t}}{\theta_{r}}=\nonumber \\
-\frac{1}{\theta_{r}}g_{r,t}+\left(\phi-1-\sum_{k=1}^{K}\alpha_{k}\right)\ln g_{r,t}-\left(\sum_{k=1}^{K}\frac{\bar{o}_{r,kt}}{b_{k,i}}\right)\frac{1}{g_{r,t}}.
\label{eq:log_likpost_gain}
\end{gather}
Eq. (\ref{eq:log_likpost_gain}) corresponds to a generalized inverse Gaussian (GIG) distribution \cite{Kawamura} with parameters $\vartheta=\phi-\sum_{k=1}^{K}\alpha_{k}$,
$\rho=\frac{1}{\theta_{r}}$, and $\tau=\sum_{k=1}^{K}\frac{\bar{\mathbf{o}}_{r,t}}{b_{k,i}}$. The GIG distribution is generally defined as:
\begin{gather*}
\ln\text{GIG}\left(g;\vartheta,\rho,\tau\right)=-\rho g+\left(\vartheta-1\right)\ln g-\frac{\tau}{g}+\\
\frac{\vartheta}{2}\ln\rho-\ln2-\frac{\vartheta}{2}\ln\tau-\ln\mathcal{K}_{\vartheta}\left(2\sqrt{\rho\tau}\right),
\end{gather*}
for $g\geq0,\rho\geq0,$ and $\tau\geq0$. Here, $\mathcal{K}_{\vartheta}\left(\cdot\right)$ denotes a modified Bessel function of the second kind. The required
expectations are given as \cite{Kawamura}:
\begin{gather}
E\left(G\right)=\frac{\mathcal{K}_{\vartheta+1}\left(2\sqrt{\rho\tau}\right)\sqrt{\tau}}{\mathcal{K}_{\vartheta}\left(2\sqrt{\rho\tau}\right)\sqrt{\rho}},\label{eq:expected_G}\\
E\left(G^{-1}\right)=\frac{\mathcal{K}_{\vartheta-1}\left(2\sqrt{\rho\tau}\right)\sqrt{\rho}}{\mathcal{K}_{\vartheta}\left(2\sqrt{\rho\tau}\right)\sqrt{\tau}},\label{eq:expected_G-1}\\
E\left(\ln G\right)=\frac{\frac{\partial\mathcal{K}_{v}\left(2\sqrt{\rho\tau}\right)}{\partial v}|_{v=\vartheta}}{\mathcal{K}_{\vartheta}\left(2\sqrt{\rho\tau}\right)}+\ln\sqrt{\frac{\tau}{\rho}}.
\label{eq:expected_lnG}
\end{gather}
The posterior distribution
of the stochastic gain variable of the noise signal can be obtained similarly.
\vspace{-4mm}
\section{Gradient and Hessian for babble states\label{sec:Gradient-and-Hessian}}
\vspace{-1mm}
The gradient of $P_{2}$ evaluated at $\widehat{\ddot{\mathbf{s}}_{i}^{\prime}}$, which is used in the CCCP procedure in Subsection \ref{sub:Noise-Training}, is simply given as:
\begin{equation}
\nabla P_{2}\left(\widehat{\ddot{\mathbf{s}}_{i}^{\prime}}\right)=\left[\frac{\partial P_{2}\left(\widehat{\ddot{\mathbf{s}}_{i}^{\prime}}\right)}{\partial\widehat{\ddot{s}_{mi}^{\prime}}}\right]=\sum_{r,t,k}\omega_{t,r}\left(\ddot{\mathbf{s}}_{i}^{\prime}\right)\mathbf{b}_{k.}^{\top}\left(\frac{\beta_{k}}{\left[\mathbf{b}\widehat{\ddot{\mathbf{s}}_{i}^{\prime}}\right]_{k}}\right),
\label{eq:gradient_CCCP}
\end{equation}
where $\mathbf{b}_{k.}$ denotes the $k^{\text{th}}$ row of the basis matrix $\mathbf{b}$, and $'\top'$ denotes the transpose. Denoting $C\left(\mathbf{x}\right)=P_{1}\left(\mathbf{x}\right)+\mathbf{x}^{\top}\nabla P_{2}(\widehat{\ddot{\mathbf{s}}_{i}^{\prime}})$,
the gradient and the hessian of the cost function in (\ref{eq:CCCP_problem})
are also given as:
\begin{gather}
\nabla C\left(\mathbf{x}\right)=\left[\frac{\partial C\left(\mathbf{x}\right)}{\partial x_{m}}\right]=\nabla P_{2}\left(\widehat{\ddot{\mathbf{s}}_{i}^{\prime}}\right)-\nonumber \\
\sum_{r,t,k}\omega_{t,r}\left(\ddot{\mathbf{s}}_{i}^{\prime}\right)\mathbf{b}_{k.}^{\top}\left(\frac{\ddot{o}_{r,kt}}{\left[\mathbf{b}\mathbf{x}\right]_{k}^{2}}E_{H_{r,t}\mid\ddot{\mathbf{S}}_{r,t},\lambda}\left(H_{r,t}^{-1}\right)\right),\label{eq:gradient_state_nois}\\
\nabla^{2}C\left(\mathbf{x}\right)=\left[\frac{\partial^{2}C\left(\mathbf{x}\right)}{\partial x_{m}\partial x_{n}}\right]=\nonumber \\
\sum_{r,t,k}\omega_{t,r}\left(\ddot{\mathbf{s}}_{i}^{\prime}\right)\mathbf{b}_{k.}^{\top}\mathbf{b}_{k.}\left(\frac{2\ddot{o}_{r,kt}}{\left[\mathbf{b}\mathbf{x}\right]_{k}^{3}}E_{H_{r,t}\mid\ddot{\mathbf{S}}_{r,t},\lambda}\left(H_{r,t}^{-1}\right)\right).
\label{eq:hessian_state_nois}
\end{gather}
\vspace{-4mm}
\section*{Acknowledgment}
Part of this work was supported by the EU Initial Training Network AUDIS (grant 2008-214699). The authors would like to thank W. Bastiaan
Kleijn for a useful discussion about the babble model.
\vspace{-4mm}
\ifCLASSOPTIONcaptionsoff
  \newpage
\fi
\bibliographystyle{IEEEtran}
{
\bibliography{refs}


\end{document}